\documentclass[reprint,twoside,superscriptaddress,floatfix]{revtex4-1}
\usepackage{natbib}

\usepackage{color}

\usepackage{graphicx,epsfig,verbatim,enumerate}
\usepackage{amssymb,amsmath,amsthm}
\usepackage{ifthen}
\usepackage{morefloats}

\usepackage{algorithm}
\usepackage{algpseudocode} 

\usepackage{longtable}
\usepackage{booktabs}
\usepackage{tabularx}
\usepackage{array}

\usepackage{hyperref}
\hypersetup{colorlinks=true, linkcolor=blue}

\usepackage{textcomp}

\usepackage{mathtools}
\usepackage{fancyhdr}
\newboolean{twocolswitch}

\newcommand{\sindex}[1]{}
\newcommand{\nindex}[1]{}

\newcommand{\www}[1]{\url{#1}}

\usepackage{lettrine}

\newtheorem{definition}{Definition}[section]

\newcommand{\sgn}{\text{sgn}}

\begin{document}

\title{Fragmentation and inefficiencies in US equity markets: Evidence from the Dow 30}

\author{Brian F. Tivnan \thanks{btivnan@mitre.org}}
\affiliation{The MITRE Corporation, McLean, VA 22102} 
\affiliation{Vermont Complex Systems Center, University of Vermont, Burlington, VT 05405}
\affiliation{Department of Mathematics and Statistics, University of Vermont, Burlington, VT 05405}
\affiliation{Computational Finance Lab, Burlington, VT 05405}
\thanks{Corresponding authors: Brian Tivnan (btivnan@mitre.org) and Christopher Danforth (chris.danforth@uvm.edu).}

\author{David Rushing Dewhurst \thanks{david.dewhurst@uvm.edu}}
\affiliation{The MITRE Corporation, McLean, VA 22102}
\affiliation{Vermont Complex Systems Center, University of Vermont, Burlington, VT 05405}
\affiliation{Computational Finance Lab, Burlington, VT 05405}
\affiliation{Department of Computer Science, University of Vermont, Burlington, VT 05405}
\affiliation{Computational Story Lab, University of Vermont, Burlington, VT 05405}

\author{Colin M. Van Oort \thanks{cvanoort@uvm.edu}}
\affiliation{The MITRE Corporation, McLean, VA 22102}
\affiliation{Vermont Complex Systems Center, University of Vermont, Burlington, VT 05405}
\affiliation{Computational Finance Lab, Burlington, VT 05405}
\affiliation{Department of Computer Science, University of Vermont, Burlington, VT 05405}
\affiliation{Computational Story Lab, University of Vermont, Burlington, VT 05405}

\author{John H. Ring IV \thanks{jhring@uvm.edu}}
\affiliation{The MITRE Corporation, McLean, VA 22102}
\affiliation{Vermont Complex Systems Center, University of Vermont, Burlington, VT 05405}
\affiliation{Computational Finance Lab, Burlington, VT 05405}
\affiliation{Department of Computer Science, University of Vermont, Burlington, VT 05405}
\affiliation{Computational Story Lab, University of Vermont, Burlington, VT 05405}

\author{Tyler J. Gray
\thanks{tyler.gray@uvm.edu}}
\affiliation{Vermont Complex Systems Center, University of Vermont, Burlington, VT 05405}
\affiliation{Department of Mathematics and Statistics, University of Vermont, Burlington, VT 05405}
\affiliation{Computational Finance Lab, Burlington, VT 05405}
\affiliation{Computational Story Lab, University of Vermont, Burlington, VT 05405}

\author{Brendan F. Tivnan}
\affiliation{Computational Finance Lab, Burlington, VT 05405}
\affiliation{School of Engineering, Tufts University, Medford, MA 02155}

\author{Matthew T. K. Koehler \thanks{mkoehler@mitre.org}}
\affiliation{The MITRE Corporation, McLean, VA 22102}
\affiliation{Computational Finance Lab, Burlington, VT 05405}

\author{Matthew T. McMahon \thanks{mmcmahon@mitre.org}}
\affiliation{The MITRE Corporation, McLean, VA 22102}
\affiliation{Computational Finance Lab, Burlington, VT 05405}

\author{David Slater \thanks{dslater@mitre.org}}
\affiliation{The MITRE Corporation, McLean, VA 22102}

\author{Jason Veneman \thanks{jveneman@mitre.org}}
\affiliation{The MITRE Corporation, McLean, VA 22102}
\affiliation{Computational Finance Lab, Burlington, VT 05405}

\author{Christopher M. Danforth \thanks{chris.danforth@uvm.edu}}
\affiliation{Vermont Complex Systems Center, University of Vermont, Burlington, VT 05405}
\affiliation{Department of Mathematics and Statistics, University of Vermont, Burlington, VT 05405}
\affiliation{Computational Finance Lab, Burlington, VT 05405}
\affiliation{Computational Story Lab, University of Vermont, Burlington, VT 05405}

\begin{abstract}
Using the most comprehensive source of commercially available data on the US National Market System, we analyze all quotes and trades associated with Dow 30 stocks in calendar year 2016 from the vantage point of a single and fixed frame of reference.
We find that inefficiencies created in part by the fragmentation of the equity marketplace are relatively common and persist for longer than what physical constraints may suggest.
Information feeds reported different prices for the same equity more than 120 million times, with almost 64 million dislocation segments featuring meaningfully longer duration and higher magnitude.
During this period, roughly 22\% of all trades occurred while the SIP and aggregated direct feeds were dislocated.
The current market configuration resulted in a realized opportunity cost totaling over \$160 million, a conservative estimate that does not take into account intra-day offsetting events.
\end{abstract}

\maketitle

\section{Introduction}
\label{sec:introduction}

\begin{table*}[!htp]
	\begin{tabular}{|c|l|r|} 
		\hline
		1  & Total Opportunity Cost         &       \$160,213,922.95   \\
		2  & SIP Opportunity Cost           &       \$122,081,126.40   \\
		3  & Direct Opportunity Cost        &        \$38,132,796.55   \\
		4  & Trades                         &         392,101,579      \\
		5  & Differing Trades               &          87,432,231      \\
		6  & Traded Value                   & \$3,858,963,034,003.48   \\
		7  & Differing Traded Value         &   \$900,535,924,961.72   \\
		8  & Fraction of differing trades   &                   0.2230 \\
		9  & Fraction of differing notional &                   0.2334 \\
		10 & Ratio of (9) over (8)          &                   1.0465 \\ \hline
	\end{tabular}
	\caption{
		The SIP feed consistently displayed worse prices than the aggregate direct feed for liquidity demanding market participants during periods of dislocation, with a \$84 million net difference in opportunity cost.
		Statistics 8 - 10 indicate that trades occurring during dislocations involve approximately 5\% more value per trade on average than those that occur while feeds are synchronized.
		The values reported above are sums of daily observations, except for statistics 8 - 10, and are conservative estimates of the true, unobserved quantities since positive (favoring the SIP) and negative (favoring the direct feeds) ROC can cancel in summary calculations.
	}
	\label{tab:eyecatchers}
\end{table*}

The Dow Jones Industrial Average, colloquially known as the Dow 30, is a group of 30 equity securities (stocks) selected by S\&P Dow Jones Indices that is intended to reflect a broad cross-segment of the US economy (all industries except for utilities and transportation) \cite{dow_official}.
The Dow 30 is one of the best known indices in the US and is broadly used as a barometer of the economy.
Thus, while the group of securities that composes the Dow 30 is in some sense an arbitrary collection, it derives economic import from its ascribed characteristics.
We study the behavior of these securities as traded in modern US equity markets, known as the National Market System (NMS).
The NMS is comprised of 13 networked exchanges coupled by information feeds of differential quality and subordinated to national regulation.
Adding another layer of complexity, the NMS supports a diverse ecosystem of market participants, ranging from small retail investors to institutional financial firms and designated market makers.

We do not attempt to unravel and attribute the activity of each of these actors here; several others have attempted to classify such activities with varying degrees of success in diverse markets \cite{kirilenko2011flash, goldstein2004trading, grinblatt2000investment}.
We take a first-principles approach by compiling an exhaustive catalog of every dislocation, 
defined as a nonzero pairwise difference between the prices displayed by the National Best Bid and Offer (NBBO), as observed via the Securities Information Processor (SIP) feed, and Direct Best Bid and Offer (DBBO), as observed via the consolidation of all direct feeds.

The SIP and consolidation of all direct feeds are representative of the displayed quotes from the national exchanges (lit market).
Additionally, we catalog every trade that occurred in the NMS among the Dow 30 in calendar year 2016, allowing an investigation of the relationship between trade execution and dislocations. 
We compile a dataset of all trades that may lead to a non-zero realized opportunity cost (ROC).
We find that dislocations---times during which best bids and offers (BBO) reported on different information feeds observed at the same time from the point of view of a unified observer differ---and differing trades---trades that occur during dislocations---occur frequently.
We measure more than 120 million dislocation segments, events derived from dislocations between the NBBO and DBBO, in the Dow 30 in 2016, summary statistics of which are displayed in Table \ref{tab:eyecatchers}.
Approximately 65 million of those dislocation segments are what we term {\it actionable}, meaning that we estimate that there exists a nontrivial likelihood that an appropriately equipped market participant could realize arbitrage profits due to the existence of such a dislocation segment.
(We discuss actionability in detail in Sec.\ \ref{subsec:physical} and the role that potential arbitrageurs play in the functioning of the NMS in Sec.\ \ref{sec:concludingremarks}.)
Market participants incurred an estimated \$160 million USD in opportunity cost due to information asymmetry between the SIP and direct feed among the Dow 30 in 2016. 
We calculate the ROC using the NBBO price as the baseline.
Deviations from this price contribute to the ROC with positive sign if the direct feed displays a worse price than the SIP, or with negative sign if the direct feed displays a better price than the SIP (from the perspective of a liquidity demanding market participant). 

To characterize these phenomena, we use a publicly available dataset that features the most comprehensive view of the NMS (see Sec.\ \ref{sec:data} below) and is effectively identical to that used by the Securities and Exchange Commission's (SEC) Market Information Data Analytics System (MIDAS).
In addition to its comprehensive nature, this data was collected from the viewpoint of a unified observer: a single and fixed frame of reference co-located from within the Nasdaq data center in Carteret, N.J.
We are unaware of any other source of public information (i.e., dataset available for purchase) or private information (e.g., available only to government agencies) that is collected using the viewpoint of a single, unified observer.

We demonstrate that the topological configuration of the NMS entails endogenous inefficiency.
The fractured nature of the auction mechanism, continuous double auction operating on 13 heterogeneous exchanges and at least 35 Alternative Trading Systems (ATSs) \cite{finra_ats_quarterly}, is a consistent generator of dislocations and opportunity cost realized by market participants.

\section{Literature Review}\label{sec:theory}
\subsection{Theory of market efficiency}
The efficient markets hypothesis (EMH) as proposed by Fama \cite{fama1970efficient} has left an indelible mark upon the theory of financial markets.
Analysis of transaction data from the late 1960s and early 1970s strongly suggested that individual equity prices, and thus equity markets, fully incorporated all relevant publicly available information - the typical definition of market efficiency.
A stronger version of the EMH proposes the incorporation of private information as well, via insider trading and other mechanisms.
Previous studies have identified exceptions to this hypothesis \cite{bouchaud2019econophysics}, such as price characteristics of equities in emerging markets \cite{foye2013persistence}, the existence of momentum in the trajectories of equity prices \cite{fama2012size}, and speculative asset bubbles.
Recent work by Fama and French has demonstrated that the EMH remains largely valid \cite{fama2012size} when price time series are examined at timescales of at least 20 minutes and over a sufficiently long period of time.
However, the NMS operates at speeds far beyond that of human cognition \cite{johnson2013abrupt} and consists of fragmented exchanges \cite{ohara2015high} that may display different prices to the market.
More permissive theories on market efficiency, such as the Adaptive Markets Hypothesis \cite{lo2004adaptive}, allow for the existence of phenomena such as dislocations due to reaction delays, faulty heuristics, and information asymmetry \cite{akerlof1978market}.
In line with this, the Grossman-Stiglitz paradox \cite{grossman1980impossibility} claims that markets cannot be perfectly efficient in reality, since market participants would have no incentive to obtain additional information.
If market participants do not have an incentive to obtain additional information, then there is no mechanism by which market efficiency can improve. 
The proposition that markets are not perfectly efficient is supported by recent research.
O'Hara \cite{ohara2015high}, Bloomfeld \cite{bloomfield2009noise}, Budish \cite{budish2015high}, and others provide evidence that well-informed traders are able to consistently beat market returns as a result of both structural advantages and the actions of less-informed traders, so called “noise traders" \cite{black1986noise}.
This compendium of results points to a synthesis of the competing viewpoints of market efficiency.
Specifically, that financial markets do seem to eventually incorporate all publicly available information, but deviations can occur at fine timescales due to market fragmentation and information asymmetries.

\subsection{Empirical studies of market dislocations}
Since the speed of information propagation is bounded above by the speed of light in a vacuum, it is not possible for information to propagate instantaneously across a fragmented market with spatially separated matching engines, such as the NMS.
These physically-imposed information propagation delays lead us to expect some decoupling of BBOs across both matching engines and information feeds.
Such divergences were found between quotes on NYSE and regional exchanges as long ago as the early 1990s \cite{blume1991differences}, in NYSE securities writ large \cite{lee1993market}, in Dow 30 securities in particular \cite{hasbrouck1995one}, between NASDAQ broker-dealers and ATSs as recently as 2008 \cite{barclay2003competition, shkilko2008locked}, and in NASDAQ listed securities as recently as 2012 \cite{ding2014slow}.
U.S.\ equities markets have changed substantially in the intervening years, hence the motivation for our research.
It is \textit{a priori} unclear to what extent dislocations should persist within the NMS beyond the round-trip time of communication via fiber-optic cable.
A first-pass analysis of latencies between matching engines could conclude that, since information traveling at the theoretical speed of light between Mahwah and Secaucus would take approximately 372 $\mu$s to make a round trip between those locations, then dislocations of this length might be relatively common.
However, a light-speed round trip between Secaucus and Mahwah takes approximately 230 $\mu$s and between Secaucus and Carteret takes approximately 174 $\mu$s.
Enterprising agents at Secaucus could rectify the differences in quotes between Mahwah and Carteret without direct interaction between agents in Carteret and agents in Mahwah.

\begin{figure*}[tp!]
\centering	
\includegraphics[width=.975\textwidth]{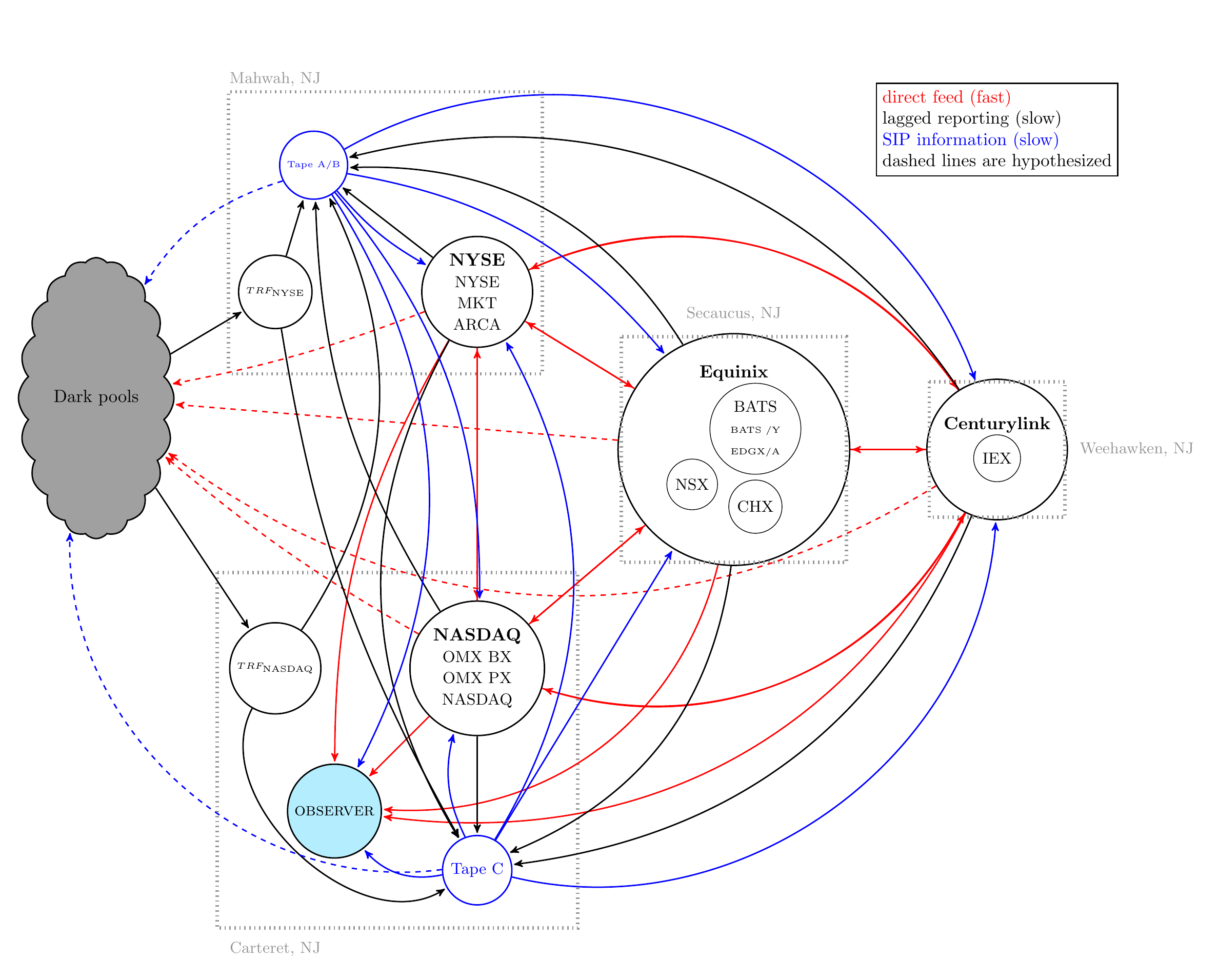} 
\caption{
The NMS (lit market and ATSs) as implied by the comprehensive market data.
As we do not have the specifications of inter-market center communication mechanisms and have minimal knowledge of intra-market center communication mechanisms, we simply classify information as having high latency, as the SIP and lagged information heading to the SIP do, or low latency, as the information on the direct feeds does.
Note the existence of the observer, located in Carteret NJ.
Without a single, fixed observer it is difficult to clock synchronization issues and introduces an unknown amount of noise into measurements of dislocations and similar phenomena.
Clock synchronization issues are avoided when using data collected from a single point of presence since all messages may be timestamped by a single clock, controlled by the observer.
}
\label{fig:NMS-detailed}
\end{figure*}

Several other authors have considered the questions of calculating and quantifying the occurrence of dislocations or dislocation-like measures.
In the aggregate, these studies conclude that price dislocations do not have a substantial effect on retail investors, as these investors tend to trade infrequently and in relatively small quantities, while conclusions differ on the effect of dislocations on investors who trade more frequently and/or in larger quantities, such as institutional investors and trading firms. 
Ding, Hanna, and Hendershot (DHH) \cite{ding2014slow} investigate dislocations between the SIP NBBO and a synthetic BBO created using direct feed data.
Their study focuses on a smaller sample, 24 securities over 16 trading days, using data collected by an observer at Secaucus, rather than Carteret, and does not incorporate activity from the NYSE exchanges.
They found that dislocations occur multiple times per second and tend to last between one and two milliseconds.
In addition, DHH find that dislocations are associated with higher prices, volatility, and trading volume.
Bartlett and McCrary \cite{bartlett2019rigged} also attempted to quantify the frequency and magnitude of dislocations.
However, Bartlett and McCrary did not use direct feed data, so the existence of dislocations was estimated using only Securities Information Processor (SIP) data, making it difficult to directly align their results to those presented here. 
A study by the TABB Group of trade execution quality on midpoint orders in ATSs also noted the existence of latency between the SIP and direct data feeds, as well as the existence of intra-direct feed latency, due to differences in exchange and ATS software and other technical capabilities \cite{alexander2015dark}.
Wah \cite{wah2016prevalent} calculated the potential arbitrage opportunities generated by latency arbitrage on the S\&P 500 in 2016 using data from the SEC's MIDAS platform \cite{sec2013midas}.
Wah's study is of particular interest as it is the only other study of which we are aware that has used comprehensive data.
Though similar in this respect, the quantities estimated in Wah's study differ substantially from those considered here.
Wah located time intervals during which the highest buy price on one exchange was higher than the lowest sell price on another exchange, termed a ``latency arbitrage opportunity'' in that work, and examined the potential profit to be made by an infinitely-fast arbitrageur taking advantage of these price discrepancies. 
This idealized arbitrageur could have captured an estimated $\$3.03$B USD in latency arbitrage among S\&P 500 tickers during 2014, which is on the same order of magnitude (on a per-ticker basis) as our approximately $\$160$M USD in realized opportunity cost among Dow 30 tickers during calendar year 2016. 

Other authors have analyzed the effect of high-frequency trading (HFT) on market microstructure, which is at least tangentially related to our current work due to its reliance on low-latency, granular timescale data and phenomena.
O'Hara \cite{ohara2015high} provides a high-level overview of the modern-day equity market and in doing so outlines the possibility of dislocation segments arising from differential information speed.
Angel \cite{angel2011equity, angel2015equity} claims that price dislocations are relatively rare occurrences, while Carrion \cite{carrion2013very} provides evidence of high-frequency trading strategies' effectiveness in modern-day equity markets via successful, intra-day market timing.
Budish \cite{budish2015high} notes that high-frequency trading firms successfully perform statistical arbitrage (e.g., pairs trading) in the equities market, and ties this phenomenon to the continuous double auction mechanism that is omnipresent in the current market structure.
Menkveld \cite{menkveld2013high} analyzed the role of HFT in market making, finding that HFT market making activity correlates negatively with long-run price movements and providing some evidence that HFT market making activity is associated with increasingly energetic price fluctuations.
Kirilenko \cite{kirilenko2011flash} provided an important classification of active trading strategies on the Chicago Mercantile Exchange E-mini futures market, which can be useful in creating statistical or agent-based models of market phenomena.
Mackintosh noted the effects of both fragmented markets and differential information on financial agents with varying motives, such as high-frequency traders and long-term investors, in a series of Knight Capital Group white papers \cite{mackintosh2014need}.
These papers provide at least three additional insights relevant to our study.
The first is a comparison of SIP and direct-feed information, noting that ``all data is stale'' since, regardless of the source (i.e., SIP or direct feed), rates of data transmission are capped at the speed of light in a vacuum as discussed above.
The second is that the SIP and the direct feeds are almost always synchronized.
That is, for U.S. large cap stocks like the Dow 30 in 2016, synchronization between the SIP and direct feeds existed for 99.99\% of the typical trading day.
Stated another way, Mackintosh observed dislocations between quotes reported on the SIP and direct feeds for 0.01\% of the trading day, or a sum total of 23 seconds distributed throughout the trading day.
The third insight from the Mackintosh papers relevant to our study reflects the significance of dislocations.
Mackintosh observed that 30\% of daily value typically traded during these dislocations.

\begin{table*}
	\centering
	\textbf{NMS Propagation Delay Estimates}
	\begin{tabular}{|lrrrr|}
		\hline 
		& Carteret-Mahwah   & Mahwah-Secaucus   & Carteret-Secaucus & Secaucus-Weehawken \\ \hline
		Straight-line Distance & 34.55 mi          & 21.31 mi          & 16.22 mi          & 2.56 mi            \\
		& 55.6 km           & 34.3 km           & 26.1 km           & 4.12 km            \\
		Light speed, one-way   & 185.75 $\mu$s     & 114.57 $\mu$s     & 87.2 $\mu$s       & 13.76 $\mu$s       \\
		Light speed, two-way   & 371.5 $\mu$s      & 229.14 $\mu$s     & 174.4 $\mu$s      & 27.52 $\mu$s       \\
		Fiber, one-way         & 272.44 $\mu$s     & 168.07 $\mu$s     & 127.89 $\mu$s     & 20.19 $\mu$s       \\
		Fiber, two-way         & 544.88 $\mu$s     & 336.14 $\mu$s     & 255.78 $\mu$s     & 40.38 $\mu$s       \\
		Hybrid laser, one-way  & -                 & -                 & 94.5 $\mu$s       & -                  \\
		Hybrid laser, two-way  & -                 & -                 & 189 $\mu$s        & -                  \\ \hline
	\end{tabular}
	\caption{ 
		The speed of light is approximated by $186,000$ mi/s (or $300,000$ km/s) and fiber propagation delays are assumed to be 4.9$\mu$s/km \cite{m2optics}.
		These propagation delays form the basis for estimates of the duration required for a dislocation segment to be considered actionable, though these figures do not account for any computing delays and thus are lower bounds for the definition of actionable.
		Data center locations, distances between data centers, and one-way hybrid laser propagation delay are obtained from Anova Technologies \cite{anova-map}.
		\label{tab:propagation}
	}
\end{table*}

For a more comprehensive review of the literature on high frequency trading and modern market microstructure more generally, we refer the reader to Goldstein et al.\ \cite{goldstein2014computerized} or Chordia et al.\ \cite{chordia2013high}.
Arnuk and Saluzzi \cite{arnuk2012broken} provide a monograph-level overview of the subject from the viewpoint of industry practitioners.

\section{Description of exchange network and data feeds}
Here we provide a brief overview of the National Market System (NMS), including a description of infrastructure components and some varieties of market participants. 
In particular, we note the information asymmetry between participants informed by the Securities Information Processor and those informed by proprietary, direct information feeds. 

\subsection{Market participants}
There are, broadly speaking, three classes of agents involved in the NMS: traders, of which there exist essentially four sub-classes (retail investors, institutional investors, brokers, and market-makers) that are not mutually exclusive; exchanges and ATSs, to which orders are routed and on which trades are executed; and regulators, which oversee trades and attempt to ensure that the behavior of other market participants abides by market regulation. 
See Appendix \ref{sec:reg-nms} for an overview of select regulations.
We note that Kirilenko \textit{et al.}\ claim the existence of six classes of traders based on technical attributes of their trading activity \cite{kirilenko2011flash}.
This classification was derived from activity in the S\&P 500 (E-mini) futures market, not the equities market, but is an established classification of trading activity. 
It is not possible to perform a similar study in the NMS since agent attribution is not publicly available \footnote{
The Consolidated Audit Trail (CAT) is an SEC initiative (SEC Rule 613) that will require such attribution to be made available \cite{cat}.
At the time of writing this framework was not yet constructed.
}.
Though the scope of this work does not encompass an analysis of various classes of financial agents, we describe some important agent archetypes in Appendix \ref{sec:market-participants}. 

\begin{figure*}
	\includegraphics[width=\textwidth]{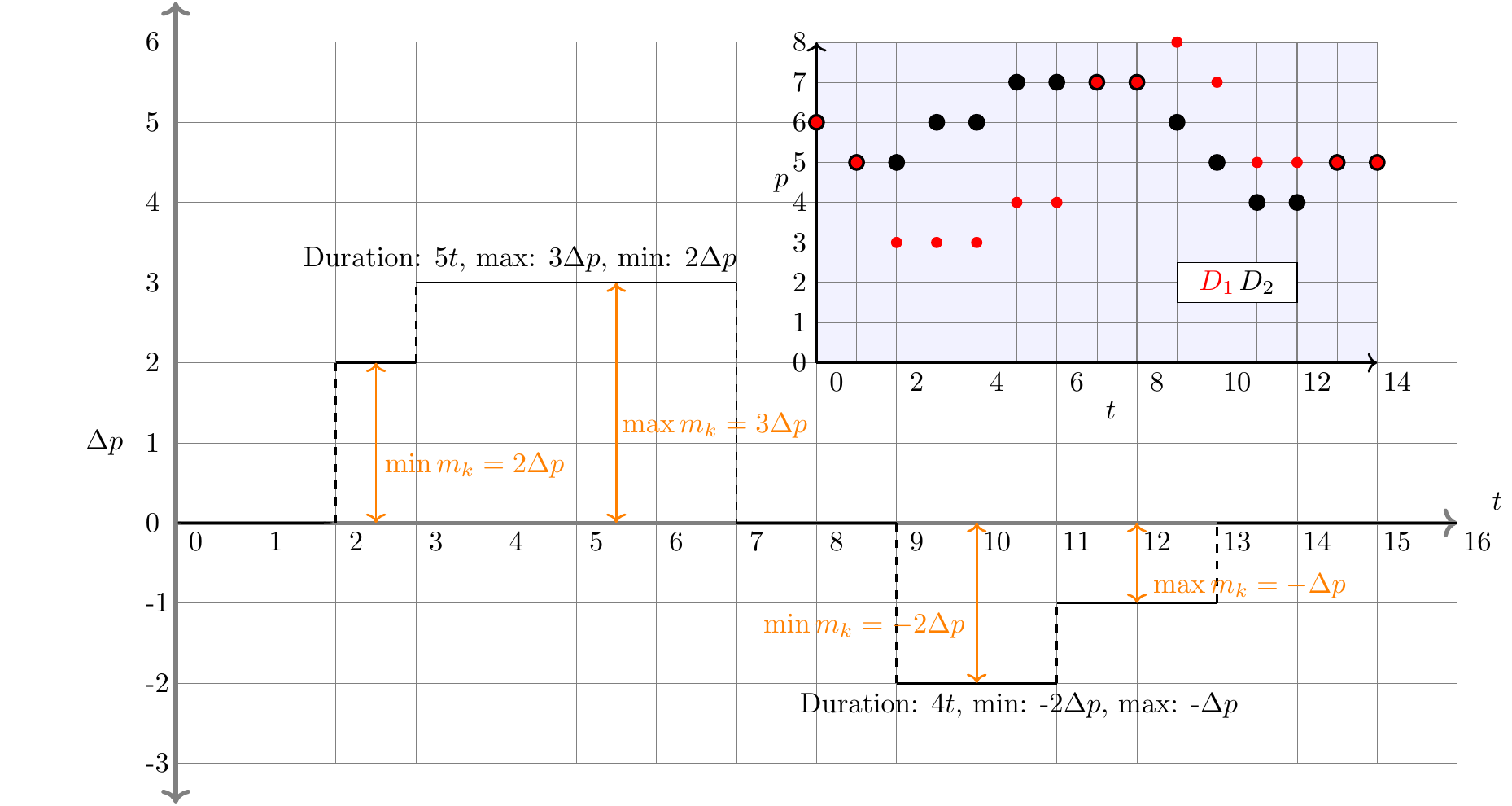}
	\caption{
		Diagram of two dislocation segments (DS).
		The inset plot shows the time series of best quotes that generate the DSs.
		Where the time series diverge from the same value, a DS occurs.
		We have deliberately not placed units on $t$, $\Delta p$, and $p$ to indicate that DSs can occur in any market in which there are differing information feeds, not just in the NMS, though we do assume that these quantities are quantized.
		In the case of the NMS, we take $t$ in units of $\mu s$ and $\Delta p$ in units of \$0.01.
		For the sake of simplicity this figure only displays one side of a hypothetical book.
		Marker size in the inset plot is used only for visual distinction.
		\label{fig:dislocation_diagram}
	}
\end{figure*}

\subsection{Physical considerations}\label{subsec:physical}
Contrary to its moniker, ``Wall Street'' is actually centered around northern New Jersey. 
The matching engines for the three NYSE exchanges are located in Mahwah, NJ, while the matching engines for the three NASDAQ exchanges are located in Carteret, NJ. 
The other major exchange families base their matching engines at the Equinix data center, located in Secaucus, NJ, except for IEX, which is based close to Secaucus in Weehawken, NJ.
The location of individual ATSs is generally not public information.
However, since there is a great incentive for ATSs to be located close to data centers (see sections \ref{sec:theory} and \ref{sec:results}), it is likely that many ATSs are located in or near the data centers that house the NMS exchanges.
For example, Goldman Sachs's Sigma X$^2$ ATS has its matching engine located at the Equinix data center in Secaucus, NJ \cite{gs_sigmax2_form_atsn}.

Since matching engines perform the work of matching buyers with sellers in the NMS, we hereafter refer to the locations of the exchanges by the geographic location of their matching engine.
For example, IEX has its point of presence in Secaucus, but its matching engine is based in Weehawken; we locate IEX at Weehawken.

\begin{figure*}
	\centering
	\includegraphics[width=\textwidth]{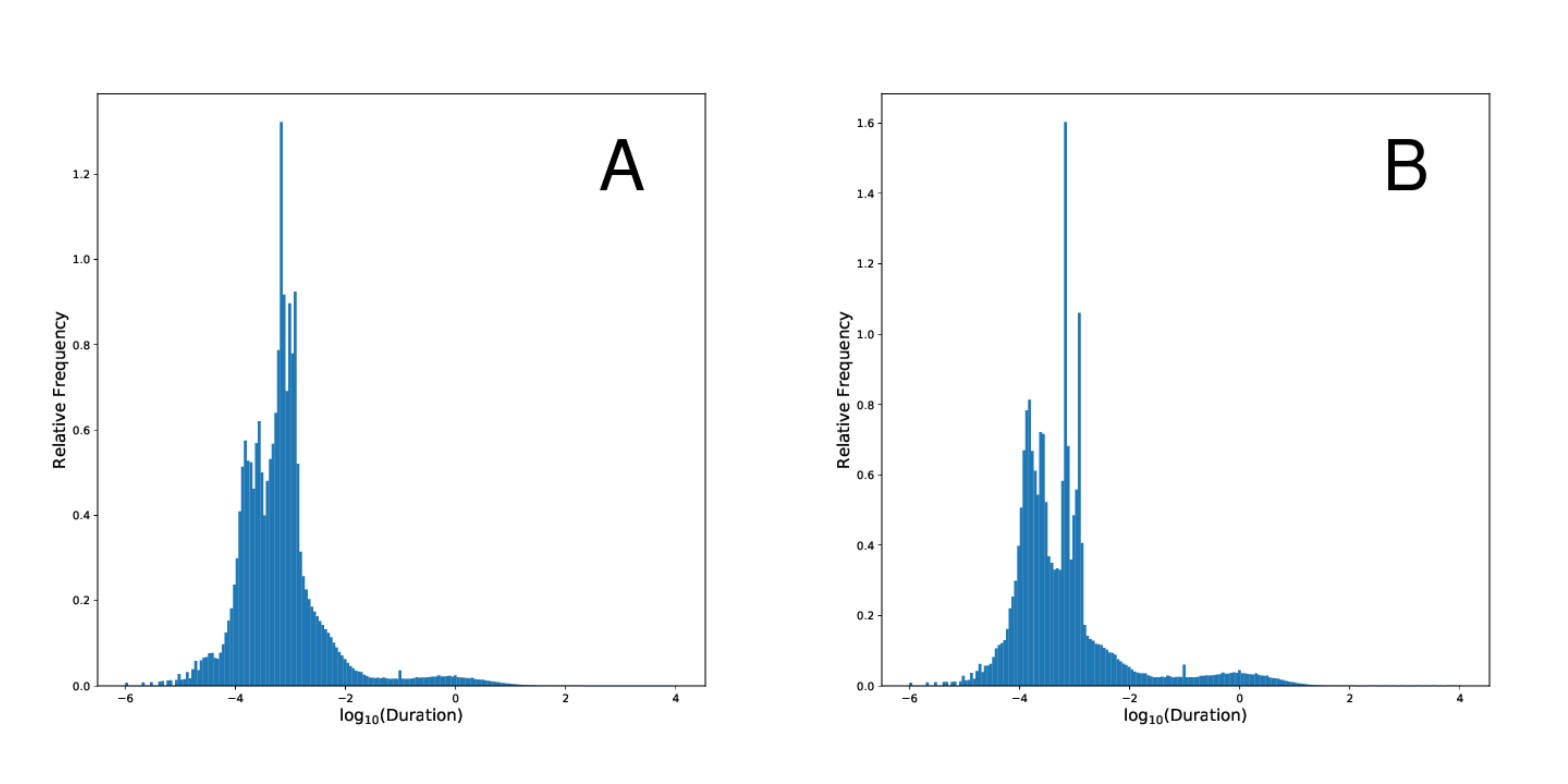}
	\caption{
		Panel A displays the distribution of dislocation segment (DS) durations.
		Panel B displays the distribution of DS durations with a magnitude greater than \$0.01.
		Both panels have a logged x-axis.
	}
	\label{fig:dur_dist}
\end{figure*}

This geographic decentralization has a profound effect on the operation of the NMS.
We calculate minimum propagation delays between exchanges and are displayed in Table \ref{tab:propagation}.
In reality, the time for a message to travel between exchanges will be strictly greater than these lower bounds, since light is slowed by transit through a fiber optic cable, and further slowed by any curvature in the cable itself.
The two-way estimates in Table \ref{tab:propagation} give a lower bound on the minimum duration required for a dislocation segment to be ``actionable" and a more realistic estimate derived by assuming propagation through a fiber optic cable with a refractive index of 1.47 \cite{m2optics}. 
These estimates do not account for computing delays, which may occur at either end of the communication lines, in order to avoid speculation.
In practice such computing delays will also have a material effect on which dislocation segments are truly actionable and will depend heavily on the performance of available computing hardware.

Connecting the exchanges are two basic types of data feeds: SIP feeds, containing quotes, trades, limit-up / limit-down (LULD) messages, and other administrative messages complied by the SIP; and direct data feeds, which contain quotes, trades, order-flow messages (add, modify, etc), and other administrative messages.
The direct data feeds operate on privately-funded and installed fiber optic cables that may have differential information transmission ability from the fiber optic cables on which the SIP data feeds are transmitted.
Latency in propagation of information on the SIP is also introduced by SIP-specific topology (SIP information must travel from a matching engine to a SIP processing node before being propagated from that node to other matching engines) and computation occurring at the SIP
processing node.
Due to the observed differential latency between the direct data feeds and the SIP data feed and the heterogeneous distance between exchanges, dislocation segments are created solely by the macro-level organization of the market system.
We note that in the intervening years since data was collected for analysis, the SIP has been upgraded substantially to lower latency arising from computation at SIP processing nodes. 

\begin{figure*}
	\centering
	\includegraphics[width=\textwidth]{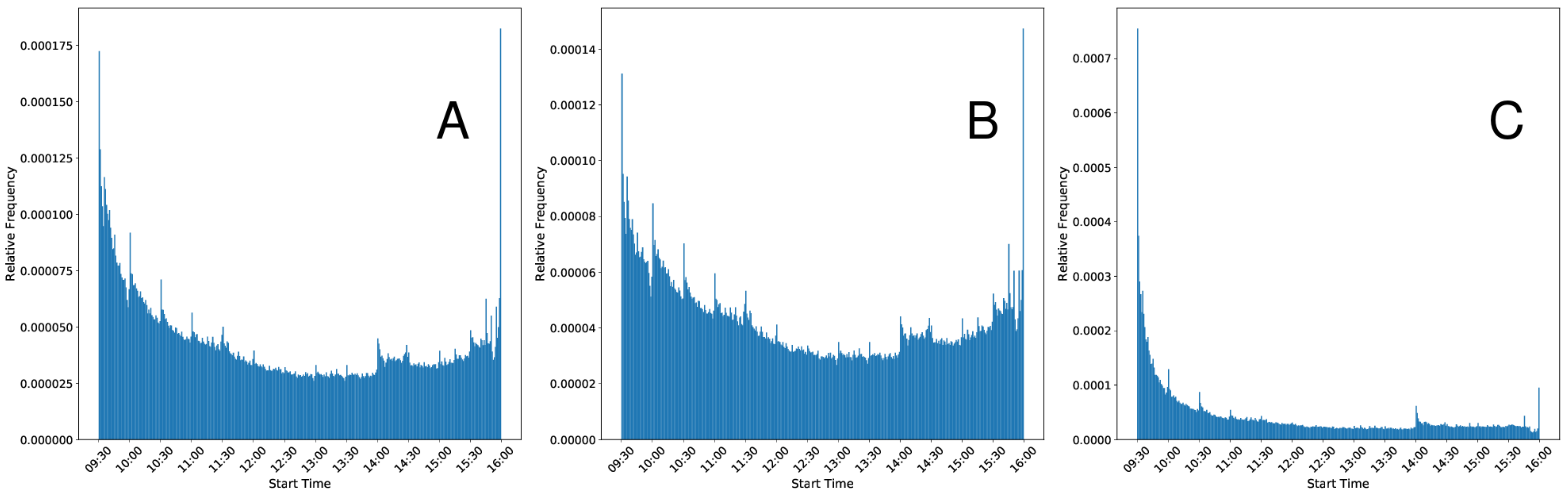}
	\caption{
		Panel A displays the distribution of dislocation segment (DS) start times binned by minute.
		Panel B displays the distribution of actionable DSs. Actionable DSs are those with a duration longer than $545 \mu s$.
		Panel C filters the actionable DSs to only include those with a minimum magnitude $> \$0.01$.
		Note that the distributions are heavily skewed right.
		A plurality of actionable DSs occur in the half-hour following the opening bell when compared to any other half-hour during the day.
		There is also a spike in the number of dislocation segments in the middle of the afternoon, which may be due to information events, such as press releases from meetings of the Federal Open Market Committee.
	}
	\label{fig:start_dist}
\end{figure*}

Our understanding of the physical layout of the NMS is depicted in Fig.\ \ref{fig:NMS-detailed} at a relatively high level. 
There are three basic types of information flow within the NMS:
\vspace{-0.8\topsep}
\begin{itemize}
\item[a.] Direct feed information, which flows to anyone who subscribes to it.
Practically speaking, direct feed information is not inexpensive (on the order of \$$130,000$ USD per month, see Table \ref{table:feed pricing} for details) and so is used primarily by exchanges, large financial firms, and ATSs. 
Direct feed information thus flows to and from the exchanges (and the major exchange participants).
We hypothesize that direct feed information also flows to ATSs, since they require some type of price signal in order for the market mechanism to function and may benefit from low latency data.
This was the case for at least one major ATS, Goldman Sachs's Sigma X$^2$, as of May 2019, so it is plausible that it is true for others \cite{gs_sigmax2_form_atsn}.
The direct feeds provide the fastest means by which to acquire a price signal, and thus may provide the best economic value to traders dependent on frequent information updates; this provides the economic foundation for our hypothesis. 
\item[b.] SIP information, which is considerably less expensive than direct feed information and exists by regulatory mandate.
However, market participants may still subscribe to the SIP as a tool for use in arbitrage; see Section \ref{sec:theory} for discussion of this possibility. 
Market participants that choose not to purchase the direct feed data might also choose to purchase the SIP data for use as a price signal and as a backup to the consolidated direct feeds.
At least one ATS, Goldman Sachs's Sigma X$^2$, uses SIP data as a backup to direct feed data and combines both data sources to construct their local BBO \cite{gs_sigmax2_form_atsn}.
\item[c.] Lagged reporting data that is not yet collated by the SIP. 
Regulation requires that exchanges report all local quote and trade activity, and that ATSs report all trade activity. 
This information is collected by the appropriate SIP tapes and then disseminated through the SIP data feeds. 
It is the responsibility of the exchanges to report their quote and trade information to the SIP, and of ATSs to report their trade information to FINRA Trade Reporting Facilities (TRF). 
Thus, though this information will be eventually visible to all subscribers to SIP or direct feed data, it differs qualitatively from that data due to its lagged nature. 

For example, suppose a trade occurs at NYSE MKT on a NASDAQ-listed security that updates the NBBO for that security. 
Since this trade occurs at Mahwah, it takes a non-negligible amount of time for the information to propagate to SIP Tape C, located in Carteret.
However, traders located at Mahwah have access to this information more quickly, possibly allowing them an information advantage over their Carteret-based competitors.
\end{itemize}

\subsection{Data}\label{sec:data}
Our study utilizes all quotes and trades associate with Dow 30 stocks that occurred in calendar year 2016 (2016-01-01 through 2016-12-31), observed via the SIP and Direct feeds from a single point of presence in Carteret, NJ.
This data is provided by Thesys Group Inc., formerly known as Tradeworx, who is the sole data provider for the SEC's MIDAS \cite{thesys2018tradeworx, sec2013midas}.
MIDAS ingests more than one billion records daily---order flow, quote updates, and trade messages---from the direct feeds of all national exchanges.
These records represent the exhaustive set of posted orders, quotes, order modifications, cancellations, trades, and administrative messages issued by national exchanges.
Prior to awarding Thesys Group the MIDAS contract \cite{fbo2012midas}, the SEC conducted a sole source selection \cite{nytimes2012midas}, thereby designating Thesys Group as the only current authoritative source for NMS data. 

\begin{table*}
	\begin{tabular}{| llrrrrrr |}
		\hline
		Filter      & Statistic &      Duration &  Min. Value &  Max. Value &   Min. Mag. &   Mean Mag. &   Max. Mag. \\
		\hline
		None        & count     & $120,355,462$ &             &             &             &             &             \\
		& mean      &      0.073712 &     -0.0012 &      0.0013 &      0.0112 &      0.0124 &      0.0137 \\
		& std       &      5.519033 &      0.1698 &      0.4815 &      0.0529 &      0.2581 &      0.5075 \\
		& min       &      0.000000 &     -141.49 &      -63.21 &        0.01 &        0.01 &        0.01 \\
		& 25\%      &      0.000216 &       -0.01 &       -0.01 &        0.01 &        0.01 &        0.01 \\
		& 50\%      &      0.000624 &        0.01 &        0.01 &        0.01 &        0.01 &        0.01 \\
		& 75\%      &      0.001190 &        0.01 &        0.01 &        0.01 &        0.01 &        0.01 \\
		& max       &     10,789.83 &      372.69 &    4,905.69 &      372.69 &    2,452.85 &    4,905.69 \\
		\hline
		Duration $> 545 \mu s$    & count     &  $65,073,196$ &             &             &             &             &             \\
		& mean      &      0.136142 &     -0.0020 &      0.0022 &      0.0109 &      0.0130 &      0.0151 \\
		& std       &      7.505197 &      0.2233 &      0.6511 &      0.0653 &      0.3474 &      0.6850 \\
		& min       &      0.000546 &     -141.49 &      -63.21 &        0.01 &        0.01 &        0.01 \\
		& 25\%      &      0.000751 &       -0.01 &       -0.01 &        0.01 &        0.01 &        0.01 \\
		& 50\%      &      0.001103 &        0.01 &        0.01 &        0.01 &        0.01 &        0.01 \\
		& 75\%      &      0.002391 &        0.01 &        0.01 &        0.01 &        0.01 &        0.01 \\
		& max       &     10,789.83 &      372.69 &    4,905.69 &      372.69 &    2,452.85 &    4,905.69 \\
		\hline
		Duration $> 545 \mu s$  & count     &   $2,872,734$ &             &             &             &             &             \\
		\&        & mean      &      0.387866 &     -0.0250 &      0.0267 &      0.0305 &      0.0564 &      0.0823 \\
		Min. Mag. $> \$0.01$ & std       &     29.566716 &      0.9046 &      1.0021 &      0.3102 &      0.7116 &      1.3115 \\
		& min       &      0.000546 &     -141.49 &      -63.21 &        0.02 &        0.02 &        0.02 \\
		& 25\%      &      0.000724 &       -0.02 &        -0.02 &        0.02 &        0.02 &        0.02 \\
		& 50\%      &      0.001207 &        0.02 &        0.02 &        0.02 &        0.02 &        0.02 \\
		& 75\%      &      0.004231 &        0.02 &        0.02 &        0.03 &        0.03 &        0.03 \\
		& max       &     10,789.83 &      372.69 &      593.43 &      372.69 &      372.84 &      593.43 \\
		\hline
	\end{tabular}
	\caption{
		Dislocation Segment (DS) attributes where the first section is unconditioned, the middle section is restricted to DSs with a duration longer than 545$\mu s$, and the final section is restricted to DSs with a duration longer than 545$\mu s$ and a minimum magnitude greater than \$0.01.
		Of the approximately 120 million DSs observed, more than 54\% of them have a duration that would allow them to be considered actionable, and about 2.4\% of them are both actionable and feature a minimum magnitude greater than \$0.01. 
		This makes the magnitude of the realized opportunity cost even more remarkable.
		Additionally, note that observed durations of ``0'' are the result of DSs that begin and end within the same microsecond, the maximum precision used for the majority of market data timestamps.
	}
	\label{tab:dislocation_summary}
\end{table*}

In addition to being the authoritative data source for the SEC's MIDAS program, another significant attribute of the Thesys data is that it is collected by a single observer from a consistent location in the NMS (the Nasdaq data center in Carteret, NJ) as depicted in Fig.\ \ref{fig:NMS-detailed}.
The single observer not only allows the user to account for the relativistic effects described above but also to directly observe dislocation segments and realized opportunity cost instead of compiling estimates of these quantities as has been done in previous studies.
At the NASDAQ data center, Thesys applies a new timestamp to each message received, including messages originating from the SIP feed or one of the direct feeds, that allows subscribers to observe information flow through the NMS in the same manner as a market participant located at the Carteret data center.
In our analysis we use this ``Thesys timestamp" to synchronize information from disparate data feeds and avoid issues that otherwise could arise from clock synchronization errors and relativistic effects.
Since this timestamp is given at the time the data arrives at the server from which the data is collected,
any discrepancies in the clocks at different exchanges, ATSs, and the SIP do not affect our measurement procedures.
This timestamping procedure is identical to that used in Ding, Hanna, and Hendershott \cite{ding2014slow}.
Ideally, we would have data from four different unified observers---an observer located at each data center---so that we could compile the different states of the market that must exist depending on physical location of observation, but we do not believe that comprehensive consolidated data is available from the point of view of observers located anywhere but at Carteret, hence our selection of this location for observation.

\section{Dislocations}
\label{sec:lat-arb}
We provide a brief definition of a dislocation segment as calculated and used in this work.
Each dislocation segment can be represented by a 4-tuple:
\begin{equation}\label{eq:dislocation}
v_n = (t_n^{\text{start}},\ t_n^{\text{end}},\ \min \Delta p,\ \max \Delta p).
\end{equation}
The maximum (resp.\ minimum) value of the dislocation segment are simply the maximum (resp.\ minimum) difference in the prices that are generating the dislocation segment over the time period $[t_n^{\text{start}}, t_n^{\text{end}})$.
The time period $[t_n^{\text{start}}, t_n^{\text{end}})$ is determined by identifying a contiguous period of time where $\Delta p > 0$ or $\Delta p < 0$.
From the above quantities the duration of the dislocation segment can also be calculated. 
The quantity $\Delta p(t)$ is the difference in the price displayed by the information feeds at time $t$ as measured and timestamped by our observer in Carteret.
From the definitions of $\max \Delta p$ and $\min \Delta p$ the reader will note that dislocation segments will tend to feature $\min ( |\min \Delta p| ) \geq \$0.01$, since the minimum tick size in the NMS is set at one penny for securities with a share price of at least \$1.00.
In collating dislocation data, we record the maximum and minimum value of each dislocation segment rather than a time-weighted average of dislocation value or other statistic for the sake of simplicity.
In much of our analysis we take the absolute values of the maximum and minimum values of each dislocation segment as the fundamental object of study as any dislocation, regardless of which feed is favored, presents an opportunity for market inefficiency.

See Fig.\ \ref{fig:dislocation_diagram} for a stylized depiction of two dislocation segments, along with annotations denoting their recorded attributes.

Based on the definition of dislocation segments given above, and fully specified in Appendix \ref{sec:definitions}, we may identify the necessary and sufficient conditions for a dislocation segment to occur.
Specifically, the market state must include two or more distinct trading locations, two or more information feeds with differing latency, and a price discrepancy.
These all follow directly from elements of the definition; such that a simple, null model configuration of a single exchange with a single data feed cannot support the existence of dislocation segments as specified here.

\section{Realized Opportunity Cost}
We used the following decision procedure to calculate realized opportunity cost:
for each trade that occurred in the NMS we checked if a price discrepancy between the SIP and consolidated direct feeds was present at the time the trade executed, from the point of view of our observer in Carteret, and counted each as a {\it differing trade}.
If the differing trade executed at a price displayed by the prevailing NBBO then a price difference was calculated, i.e. $p_{\text{SIP}} - p_{\text{direct}}$ if the liquidity-demanding order was a offer and $p_{\text{direct}} - p_{\text{SIP}}$ if the liquidity-demanding order was a bid, and 
a cost, termed the realized opportunity cost (ROC), was assigned to the trade using the number of shares multiplied by the price difference.
Depth of book was not taken into account in this calculation.
The sum total of all ROC occurrences over a day was calculated and recorded.
With this construction, positive opportunity costs indicate an incentive for liquidity demanding market participants to use the SIP feed while negative opportunity costs indicate an incentive to use the aggregated direct feeds.
By ignoring the sign of the opportunity costs, and thus which feed is favored, an aggregate or total realized opportunity cost is constructed.
Intra-day events can offset---e.g., a trade that resulted in ROC that disadvantaged direct data users and a trade that resulted in ROC that disadvantaged SIP data users could both occur on the same day, partially offsetting the total ROC due to opposite signs.
Precise definitions of quantities described here are located in Appendix \ref{sec:inefficiencies}.

As above, we provide a brief toy example of how realized opportunity cost can arise and a description of its' calculation.
A minimal example involves two traders, each of which is in the market to buy the security XYZ.
One trader places orders using the SIP NBO to determine the appropriate limit price and the other places orders using the best offer from a direct feed.
If a trade for 100 shares of XYZ executes at \$100.00 per share, the current direct best offer, when the NBBO was a SIP quote of \$100.01 per share, a trader placing a bid informed by the SIP could receive an execution that resulted in a realized opportunity cost of \$0.01 per share, or \$1.00 in total. 
Because this opportunity cost favored the direct feed, this portion of ROC would be assigned a negative value.
If, during another trade on the same day, another trade for 100 shares of XYZ executes when the direct best offer price is \$101.02 and the SIP NBO price is \$101.00 per share, the trader who places orders informed exclusively by the direct feeds could have experienced a realized opportunity cost of \$0.02 per share, or \$2.00 in total, assuming that they may have been able to find counter-parties at the SIP NBO. 
This ROC is assigned a positive value because it favors the SIP feed.
Summing these two together produces a net ROC of \$1.00, hence the conservative nature of our estimates.
If, instead, our calculation summed the absolute value of each ROC-generating event, the figure above would instead be \$3.00.
A more detailed example of ROC calculation from real trade data is located in Appendix \ref{sec:calc-roc}.
\section{Results}
\label{sec:results}

\subsection{Dislocations and dislocation segments}
\label{sec:dislocations}
We find that dislocations and dislocation segments are widespread, from the point of view of our observer in Carteret, and may have qualitative welfare effects on NMS participants, particularly large investors or investors that interact with the NMS directly on a frequent basis.
There were a total of 120,355,462 dislocation segments among Dow 30 stocks in 2016. Now, let’s assume a uniform distribution of dislocations throughout the trading day. On average, we therefore expect $\frac{120,355,462}{252 \times 6.5 \times 60^2} \approx 20.4$ dislocation segments per second.
When restricting our attention to what we term \textit{actionable} dislocation segments (those with a duration longer than 545 $\mu s$), we find that there were 65,073,196 actionable dislocation segments, or on average, $\frac{65,073,196}{252 \times 6.5 \times 60^2} \approx 11$ actionable dislocation segments every second.
Even when inspecting actionable dislocation segments with a minimum magnitude greater than 1 cent, we find that there were 2,872,734 instances of these dislocation segments, or on average, $\frac{2,872,734}{252 \times 6.5 \times 60^2} \approx 0.49$ dislocation segments per second, or almost one large and actionable dislocation segment every two seconds. 

We focus much of our subsequent analysis on the dislocation segment distribution conditioned on both duration ($> 545\mu$s) and magnitude ($>$ \$0.01)
From an academic point of view, dislocations with a minimum magnitude greater than one cent are more interesting, since one might expect many dislocations to feature a magnitude that corresponds with the price quantization---minimum tick size (\$0.01 in this case).
There are several aspects of this conditional distribution that bear special notice. 
First, the distribution of each attribute is exceptionally heavy-tailed.
In absolute value, the 75\%-iles of the minimum and maximum magnitude are three cents---but the mean in absolute value of the minimum magnitude (resp.\ maximum magnitude) is 3.05 (resp. 8.23) cents.
A similar phenomena is true for the duration distribution, displayed in Fig.\ \ref{fig:dur_dist}, where the 75\%-ile is 4231 $\mu s$, while the mean is an astounding 0.389 \textit{seconds}, almost two orders of magnitude longer.
The max magnitude, min magnitude, and duration distributions are all highly skewed, while the distributions of the maximum and minimum magnitudes are nearly identical. 
Further summary statistics on dislocations with various conditioning are displayed in Table \ref{tab:dislocation_summary}. 

Fig.\ \ref{fig:start_dist} shows the distribution of dislocation segments modulo day, binned by minute. 
Intra-day dislocation segment distributions are markedly nonuniform, with a majority of the probability mass concentrated toward the beginning of the trading day.
There is also a notable spike in the number of dislocation segments occurring in mid-afternoon and at the very end of the trading day.
Additionally, there seems to be a decaying cyclic pattern in the distribution, with spikes occurring with a 30 minute frequency.

We postulate that the mid-afternoon spike, which occurs at approximately 2:00pm, is associated with meetings of the Federal Open Market Committee (FOMC).
These meetings release economically important information such as decisions regarding federal rate changes and economic forecasts, and their impact has been noted by several market participants, including analysts at NYSE \cite{nyse_blog_fomc_1, nyse_blog_fomc_2}.
Note that the NYSE analysis of the impact of FOMC meetings is based upon a quote volatility measure, which is conceptually quite similar to the dislocations discussed in our work.
Regarding the cyclic pattern, it seems that most of this activity can be attributed to the aggregated effect of seemingly random market events.
Investigating the data without aggregation reveals that almost no days exhibit this cyclic behavior for DS occurrence, though there are many days that seem to have one or more abnormal spikes in DS occurrence at seemingly random times.
During aggregation, these potentially large spikes are not entirely smoothed out, leading to the cyclic pattern observed in Fig.\ \ref{fig:start_dist}.
Interested readers may investigate the dislocation segment occurrence distributions without aggregation by using the interactive application provided in our GitLab repository \cite{dislocation_visualizer}.

\begin{figure*}[!htb]
\centering
\includegraphics[width=0.7\textwidth]{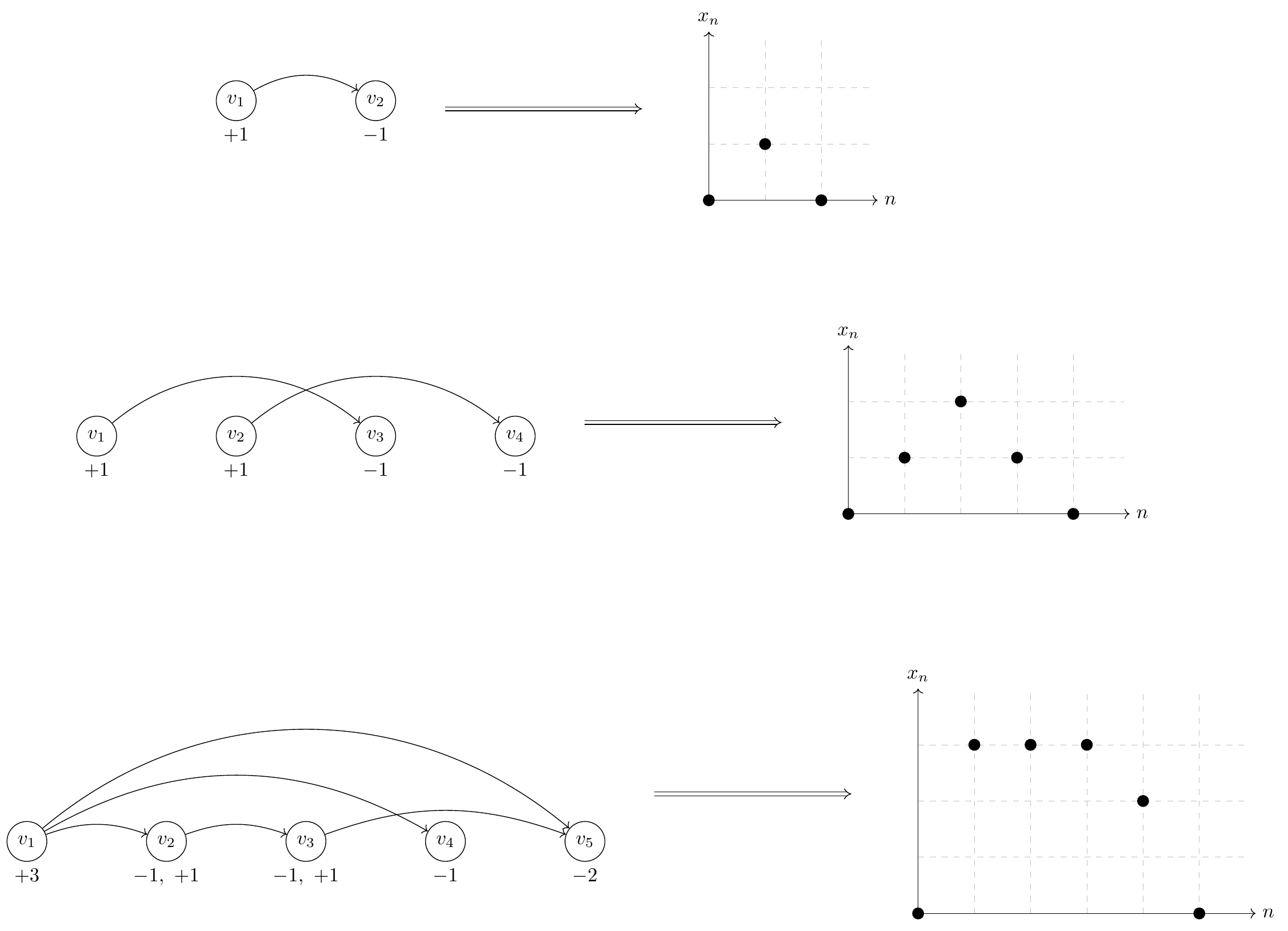}
\caption{
	A depiction of the injection mapping from an $N$-component in a ordered network to a tied positive random walk of length $N + 1$.
	The injection is given by $j \text{ outgoing edges } \cong j \text{ steps up}$ and likewise $k \text{ incoming edges } \cong k \text{ steps down}$.
	The total number of steps up or down is given by $x_{n+1} - x_n = \text{ \# of steps up $+$ \# of steps down}$.
	The top row displays a simple $2$-component, where an equity begins a dislocation at time $t_i$ and ends it at time $t_{i+1}$.
	The corresponding walk on the line starts at zero, moves up a step, and then moves down. 
	The second row displays a $4$-component identical to that described in the text of the article.
	This $4$-component, which is separable into two disconnected pieces, demonstrates the geometric nature of the ordered network.
	Since an ordering is imposed on the nodes, the crossing of the edges implies the staggered starts and stops of the two dislocations.
}
\label{fig:circle-injection}
\end{figure*}

\begin{figure*}[!htp]
\centering
\includegraphics[width=\textwidth, trim={5in 6in 7.25in 6.05in}, clip]{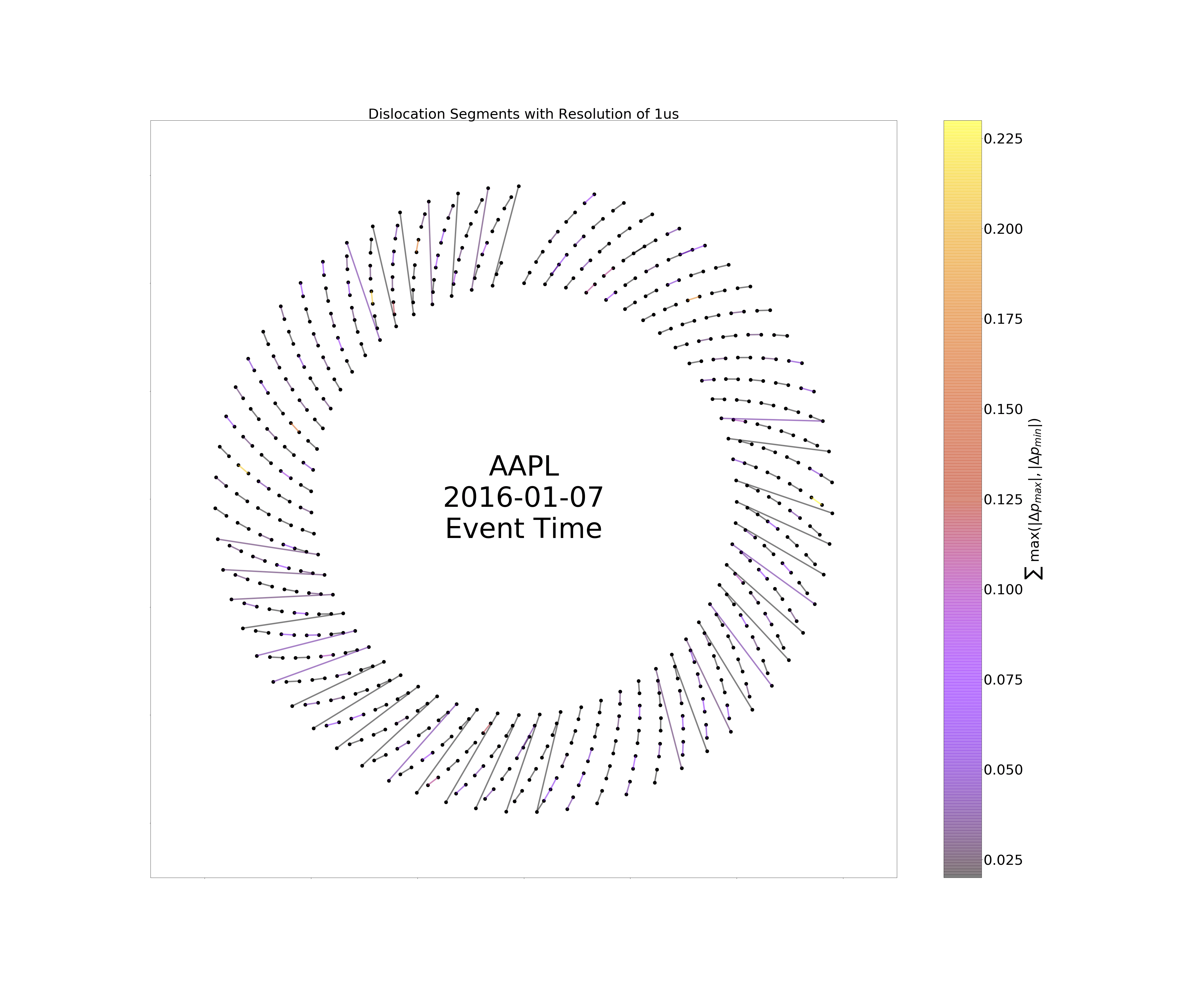}
\caption{
	Distribution of dislocation segments (DS) with minimum magnitude greater than \$0.01 and duration longer than 545$\mu s$ for AAPL on 2016-01-07 visualized with a time re-normalization procedure.
	Nodes are placed in rings modulo 10; nodes zero through 9 are in the first ray from the origin, then the angle in the plot is incremented and nodes 10 through 19 are in the second ray, etc.
	A link $e_{ij}$ connects two nodes, $v_i$ and $v_j$, if a dislocation segment starts at $v_i$ and stops at $v_j$.
	This view of the dislocation segment network preserves time ordering while defining a nonlinear transformation between uniform time ordering, as shown below in Fig.\ \ref{fig:circle-aapl-1day-real}, and uniform event-space ordering, as shown here. 
	As noted in the text, it is not necessary for only one dislocation segment to exist at the same point in time $t$. 
	For example, there are many instances of new dislocation segments starting while another is still ongoing---the first starts at $v_i$ and then another starts at $v_j$ and ends at $v_k$, followed by the first dislocation segment ending at $v_{\ell}$. 
	Irregular behavior such as this generates the banding of the edge distribution.
	Interested readers may wish to have some more context for the selected date.
	For AAPL, 2016-01-07 ranked 8th out of 252 trading days when considering ROC.
	\$106,990.23 in ROC was accumulated, which lies between the minimum of \$2,773.35 and the maximum of \$138,331.08.
	This day of AAPL also ranked 15th when considering the number of DSs.
	A total of 108,843 occurred, falling between the minimum of 9,256 and the maximum of 188,656.
}
\label{fig:circle-aapl-1day-event}
\end{figure*}

\begin{figure*}[!htp]
\centering
\includegraphics[width=\textwidth, trim={5in 6in 7.25in 6.05in}, clip]{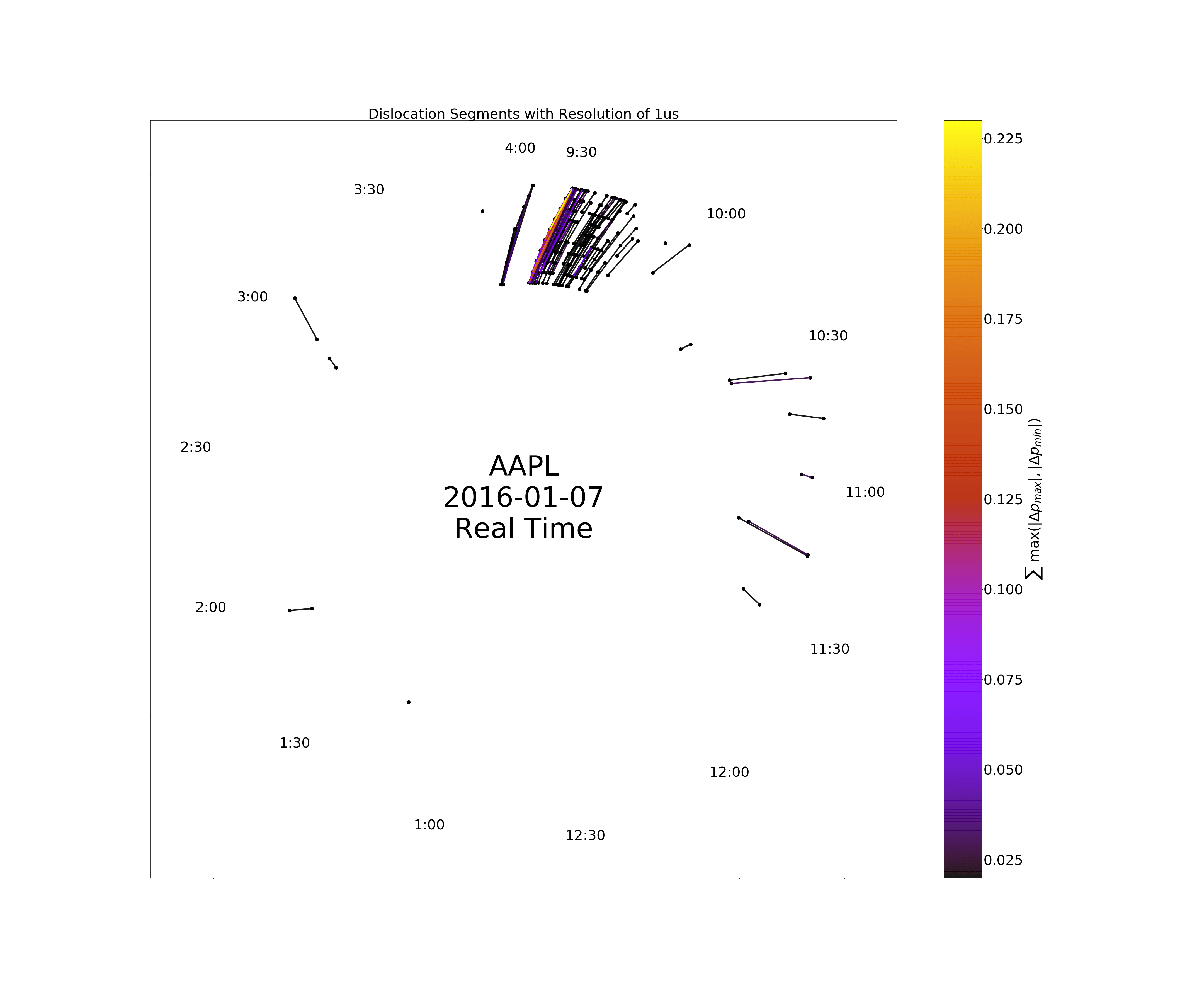}
\caption{
Dislocation segments in AAPL on 2016-01-07 without time re-normalization. 
The characteristic structure in the occurrence of dislocations segments is clearly displayed, with the majority occurring near the beginning of the trading day.
}
\label{fig:circle-aapl-1day-real}
\end{figure*}

\begin{figure*}[!htp]
\centering
\includegraphics[width=\textwidth, trim={5in 6in 7.25in 6.05in}, clip]{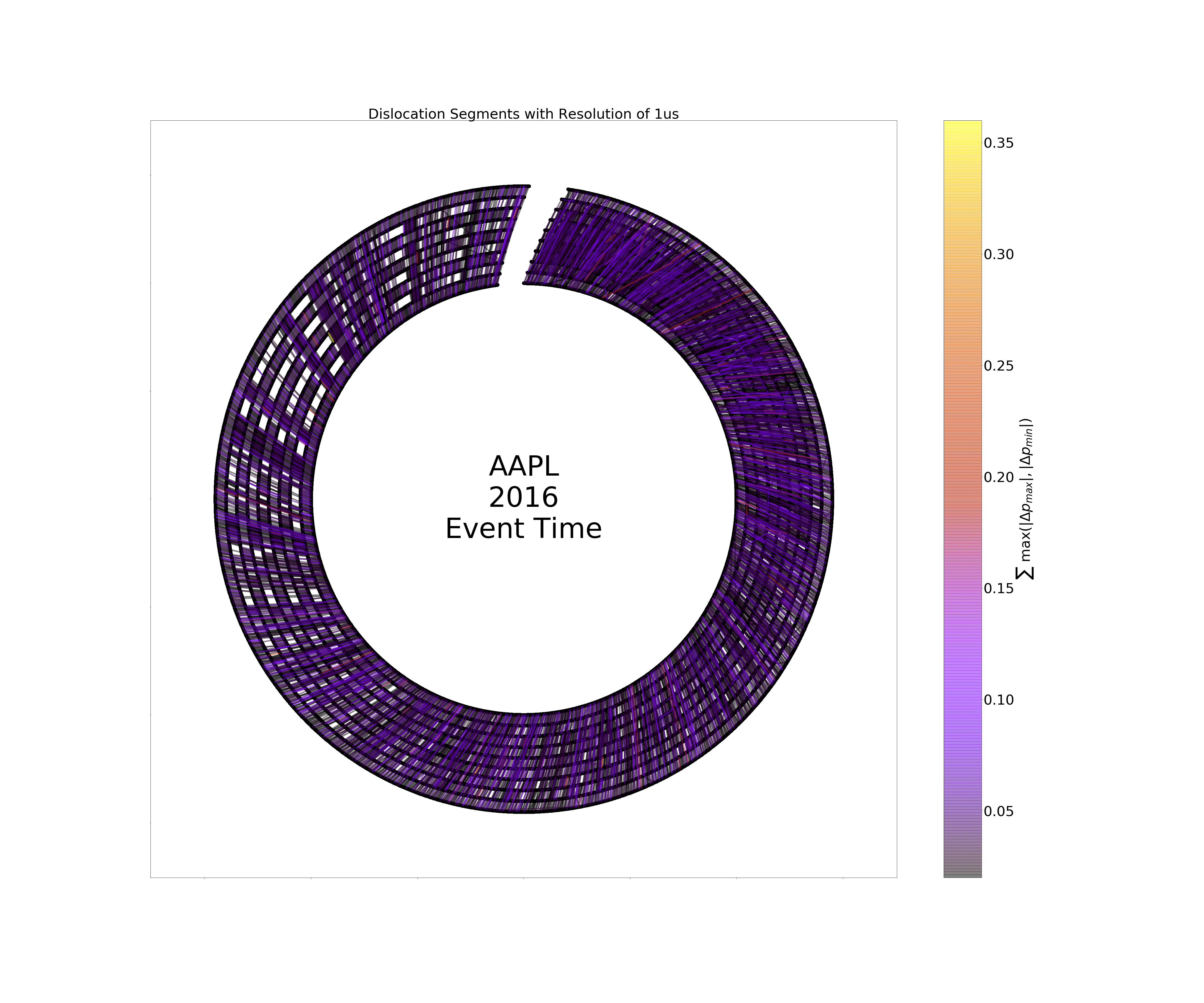}
\caption{
Dislocation segments (DS) aggregated over an entire year (modulo trading day).
Investigating structures at the trading day timescale is of interest as this is likely the longest timescale over which HFT strategies are used.
Here DSs are plotted in event space, where density is uniform between events.
Note the presence of irregular structure even here, evidence of higher-order structure in the ordering of starts and stops of DSs.
}
\label{fig:circle}
\end{figure*}

\begin{figure*}[!htp]
\centering
\includegraphics[width=\textwidth, trim={5in 6in 7.25in 6.05in}, clip]{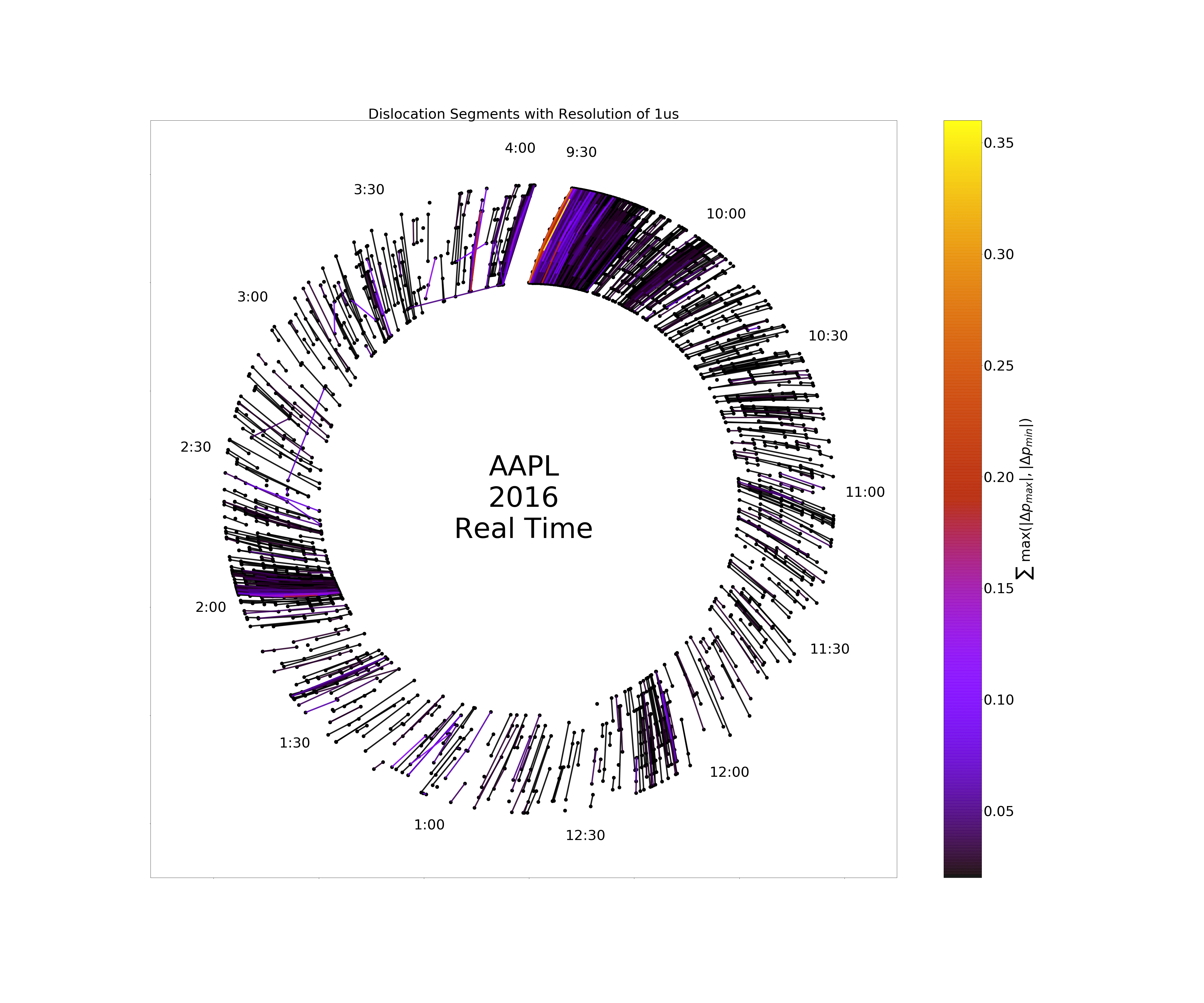}
\caption{
Dislocation segments (DS) aggregated over an entire year (modulo trading day), as above, but not transformed to event space.
The high density of dislocation segments at the beginning of the trading day, near 12:10, and near 2:00 is readily apparent.
}
\label{fig:circle-time-ordered}
\end{figure*}

To further unpack the relationship between time of day, length, and magnitude of dislocation segments, we created a representation of dislocation segments modulo day as an ordered network, termed a circle plot.  Fig.\ \ref{fig:circle-injection} illustrates the construction of the circle plots from a few toy examples. Fig.\ \ref{fig:circle-aapl-1day-event} and Fig.\ \ref{fig:circle-aapl-1day-real} depict circle plots for AAPL for an arbitrary day, whereas Fig.\ \ref{fig:circle} and Fig.\ \ref{fig:circle-time-ordered} depict circle plots for AAPL for the entirety of 2016.

Circle plots are constructed using the following algorithm. 
Starts and stops of dislocation segments at time $t$ (as measured and timestamped by our observer in Carteret) are termed events $v(t)$ and denoted by black nodes. 
More than one event can occur at each time $t$; all events are represented by the same node. 
Events $v_i(t)$ and $v_j(s)$ where $t < s$ are connected by an edge $e_{ij}$ when a dislocation segment starts at $v_i(t)$ and ends at $v_j(s)$. 
It is not necessarily the case that dislocation segments start and stop in order as seen above;
for example consider two dislocation segments, the first starting at $v_i$, and the second starting at $v_j$. The first dislocation segment could end at $v_k$, and the second could end at $v_{\ell}$. 
When $N$ events occur ``out of order'' in this way, we identify the events as a single component (even though, as in the above example, the component decomposes into two two-tuples of events) and term it an $N$-component for reasons we state below; the above example is a $4$-component.
Nodes are plotted in rays that spread outward from the geometric center of the plot in a modulo 10 relation. 
Edges between nodes $v_i$ and $v_j$ are weighted according to the quantity
\begin{equation}
\sum_{(v_i, v_j)} \max( |\Delta p_{\max}|, |\Delta p_{\min}|),
\end{equation}
where the sum is taken over all events that started at node $v_i$ and ended at node $v_j$ and $\Delta p_{\max}$ and $\Delta p_{\min}$ are the largest positive (resp.\ smallest negative) change in value that occurred during each event.
Fig.\ \ref{fig:circle-time-ordered} displays the ordered network for AAPL aggregated (modulo day) over the entire trading year.
There is high event density near the beginning of the day and there is another spike in density near noon-12:30 PM.
This clustering can make interpretation of the fine event structure difficult to discern, so we conduct a re-normalization into event space with a simple method:
consecutive events $v_i(t)$ and $v_j(s)$ are plotted in order, but at a uniform distance so that the measure on the graph becomes a Stieltjes-type instead of a Lebesgue-type measure. 
In other words, in the case of the real time representation, an event represented by a node on a fixed but arbitrary circle of the graph occurred at a multiple of 10$\mu s$ from all other events represented by nodes on the ring; 
in the case of the event-time representation, an event represented by a node on a fixed but arbitrary circle of the graph and another event represented by a node on the same circle are separated by an integer multiple of events that occurred between them.
Fig.\ \ref{fig:circle-aapl-1day-event} displays the ordered network in this re-normalized space, where it is easier to see that the usual behavior of dislocation segments is a regular cyclic, on-off (start-stop) pattern. 
However, there are multiple deviations from this pattern---any component other than a $2$-component is structurally different from a purely sequential pattern. 
In fact, there is an injection from an $N$-component and a tied, non-negative sequence $\{x_n\}_{n=0}^{N}$, $x_0 = x_{N+1} = 0$, $x_n \geq 0$ for all $n$.
This injection is defined by the relationships ``start of $k$ events $\cong$ $k$ steps up" and ``end of $k$ events $\cong$ $k$ steps down".

As a concrete example, the $4$-component described above maps to the sequence steps $\{1, 1, -1, -1\}$, with values $x_0 = 0,\ x_1 = 1,\ x_2 = 2,\ x_3 = 1,\ x_4 = 0$.
Fig.\ \ref{fig:circle-injection} displays a toy example of the injection between $N$-components in an ordered network and a tied positive sequence, as outlined above.

When aggregated over all trading days, evidence of persistent nontrivial structure in the event-space density of $N$-tuples emerges.
As stated above, Fig.\ \ref{fig:circle} and Fig.\ \ref{fig:circle-time-ordered} display the aggregate of events in AAPL modulo day. 
Visualizations of all Dow 30 securities in this format are at the authors' webpage \footnote{\href{https://compfi.org}{https://compfi.org}}.

\subsection{Realized opportunity cost}\label{sec:roc}
The large number of actionable dislocation segments likely has a direct effect on the opportunity cost market participants may incur by using one information source over the other. 
The aggregate of this realized opportunity cost can be estimated by cataloging the quantity and characteristics (average price difference, etc.) of differing trades.
Table \ref{tab:eyecatchers} summarizes many of these findings.
In the time period studied (01-01-2016 through 31-12-2016) there were a total of 392,101,579 trades of stocks in the Dow 30, with a traded value of \$3,858,963,034,003.48 USD.
Of those trades, we classified 87,432,231 trades, or 22.3\% of the total number of trades, as \textit{differing} trades, defined as follows: if the trade is on the buy side, it is a differing trade if the SIP bid is not equal to the direct bid; if the trade is on the sell side, it is a differing trade if the SIP offer is not equal to the direct offer.
These differing trades had a traded value of \$900,535,924,961.72 USD, or 23.34\% of the total traded value.
More optimal use of information presented by the SIP and direct feeds could have saved market participants a total of \$160,213,922.95 USD in ROC.
This opportunity cost was distributed unevenly, with traders informed by NBBO prices suffering \$122,081,126.40 USD in ROC, while traders informed by DBBO prices only accumulated \$38,132,796.55 USD in ROC.
Fig.\ \ref{fig:purse_by_day} displays the daily net opportunity cost aggregated over all tickers in our sample, while Table \ref{tab:roc-stats} displays realized opportunity cost summary statistics for all tickers. 
Though our observer was located in Carteret while many securities (all but four during 2016) in the Dow 30 are listed on NYSE, located in Mahwah, consultation with Table \ref{tab:roc-stats} demonstrates that mean ROC per ticker does not differ significantly by listing venue (one-way ANOVA: $F(4, 20) = 1.35$, $p = 0.25$; Kruskal-Wallis $H$-test: $H = 0.84$, $p = 0.35$).

\begin{table*}
\centering
\begin{tabular}{| lrrrrrr |}
\hline
{}   &       Trades &      Traded Value &  Diff. Trades &  Diff. Traded Value &          ROC &  ROC/Share \\
\hline
mean & 1,555,958.65 & 15,313,345,373.03 &    346,953.30 &    3,573,555,257.78 &   635,769.54 &   0.011804 \\
std  &   463,558.93 &  3,891,299,900.31 &    146,677.85 &    1,234,882,079.43 &   655,911.15 &   0.008592 \\
min  &      579,206 &  6,664,671,053.15 &        89,564 &    1,035,855,029.71 &   145,205.65 &   0.008848 \\
25\% & 1,278,813.25 & 12,915,031,172.08 &       262,209 &    2,804,569,367.64 &   417,485.73 &   0.009613 \\
50\% &    1,429,062 & 14,431,597,662.02 &       309,158 &    3,274,390,601.60 &   514,856.64 &   0.010154 \\
75\% & 1,715,351.25 & 16,829,521,684.38 &       387,772 &    3,993,470,514.97 &   666,268.27 &   0.011213 \\
max  &    3,596,006 & 30,999,914,293.66 &     1,073,029 &    9,428,952,387.10 & 7,817,684.58 &   0.098303 \\
\hline
\end{tabular}
\caption{
Summary statistics of realized opportunity cost and related statistics for Dow 30 stocks, aggregated over the 252 trading days in 2016.
}
\label{tab:purse-results}
\end{table*}

\begin{figure*}
\centering
\includegraphics[width=\columnwidth, trim={0.7cm 1.75cm 3.5cm 2cm}, clip]{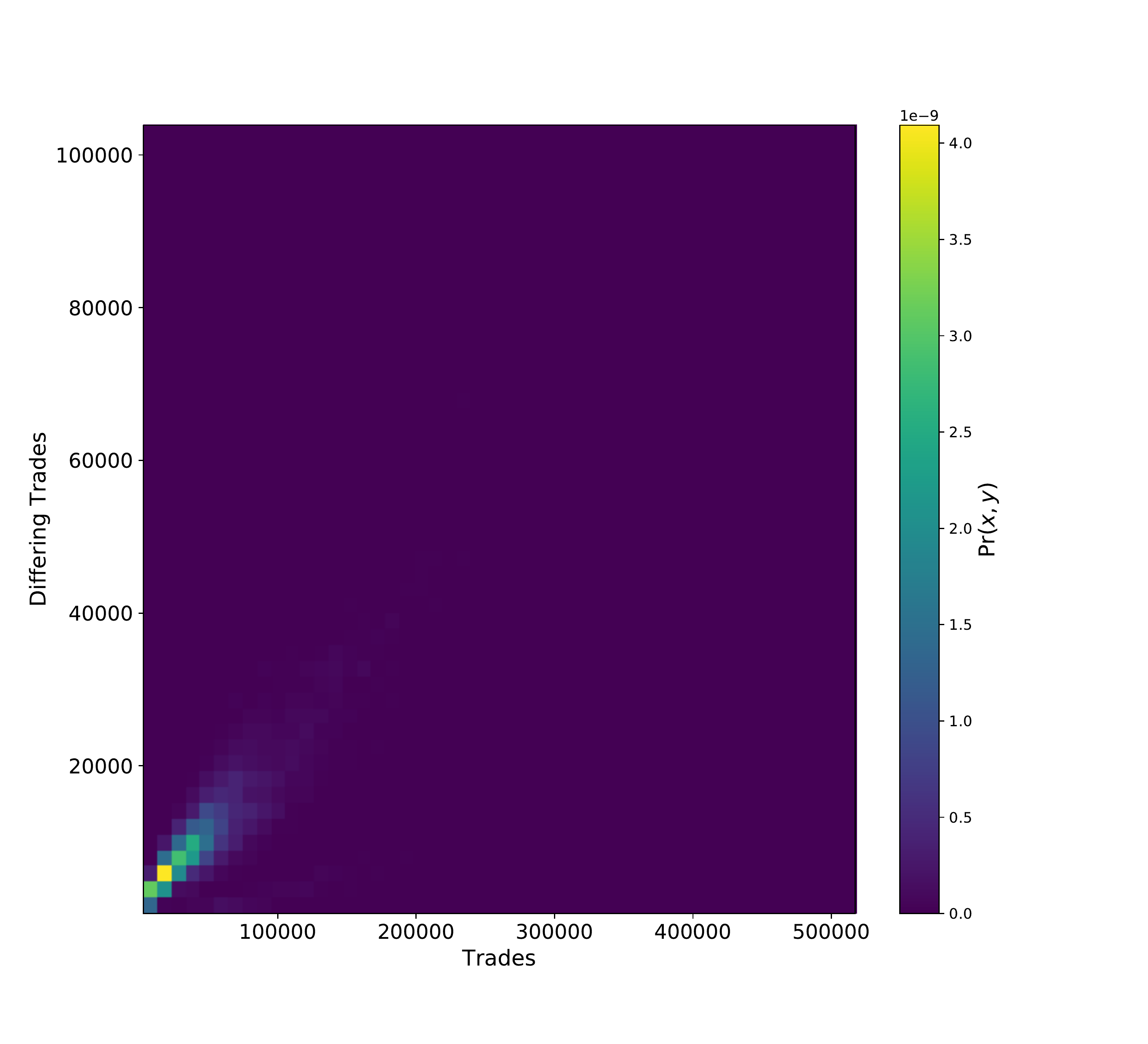}
\includegraphics[width=\columnwidth, trim={1cm 1.75cm 3.5cm 2cm}, clip]{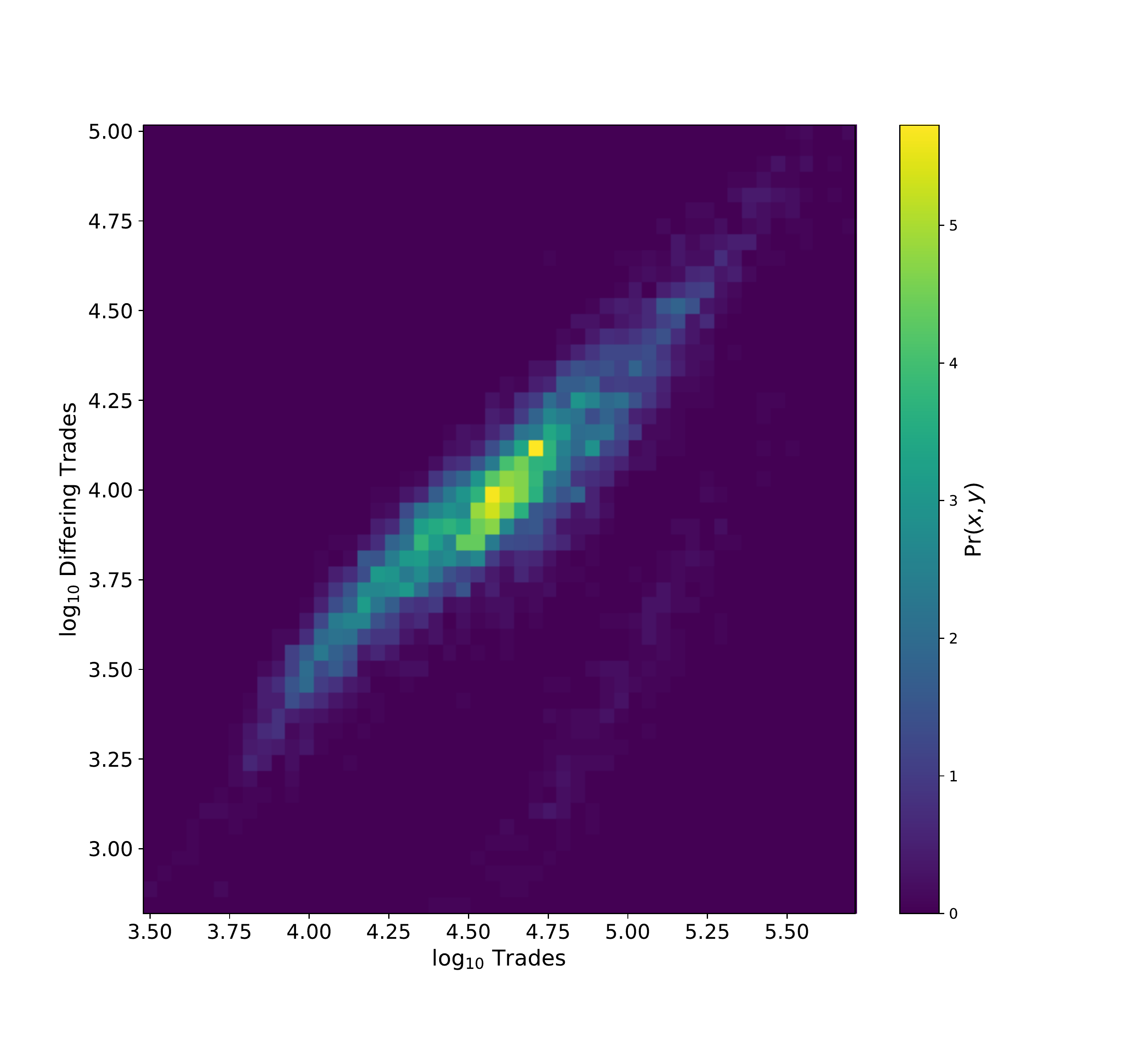}
\caption{
Left: A bivariate empirical distribution function for total trades and number of differing trades. 
Right: The same distribution, but with logged axes. 
We might expect \textit{a priori} that they are related by a constant proportion and hence should observe a fit $\log_{10}$ total trades $= c + \log_{10}$ differing trades, where $c < 0$. 
Though there is good evidence of this linear relationship, we see there is a non-negligible area of higher total trades with markedly sub-linear scaling of differing trades. 
}
\label{fig:total_vs_differing}
\end{figure*}

Fig.\ \ref{fig:total_vs_differing} provides further insight into the joint distribution of total and differing trades.
While we might {\it a priori} expect that the ratio of total to differing trades would remain roughly a fixed constant, we see that this is not observed empirically.

\section{Concluding remarks}
\label{sec:concludingremarks}
Using the most comprehensive set of NMS data publicly available, we have shown that market inefficiencies in the form of dislocations and realized opportunity cost were common in the Dow 30 in 2016 as measured by our observer in the NASDAQ data center in Carteret, NJ.
We find that inefficiencies due to the physical fragmentation of the market are widespread, totaling over \$160M USD in realized opportunity cost and 2,872,734 dislocations of magnitude $>$ \$0.01 and duration $>$ 545 $\mu$s.
These figures correspond well with those reported in other bodies of work \cite{ding2014slow, wah2016prevalent}.
Additionally, we found that the average trade that occurred during a dislocation moved approximately 5\% more value than the average trade that occurred when the NBBO and DBBO were synchronized (see Table \ref{tab:eyecatchers} row 10).
In the fifth Need for Speed report \cite{mackintosh2014need}, Mackintosh and Chen indicate that 29\% of traded value executes within a small window around quote changes, closely aligning with rows 8 and 9 from Table \ref{tab:eyecatchers}.
This may indicate that market participants could be more heavily impacted by the existence of dislocation segments than previous analyses suggest.

Though our work is empirical, our results do have implications for theoretical results on nuances of financial market efficiency. 
The discovery of systematically-different prices as measured in geographically-distinct locations that can be routinely observed by agents with access to higher-speed information flows---and cannot be routinely observed by agents without this access---has a logical bearing on questions of distributional effects of asymmetric information and market design.
This feature of fragmented market structure can be viewed as a modern-day example of the Grossman-Stiglitz paradox \cite{grossman1980impossibility}. 
Trading agents who are able to act at higher speeds may be rewarded for their investment, effort, and risk-taking behavior by executing on trading opportunities that exist for very short time intervals.
In fact, without competition among traders to reduce processing time and infrastructure providers to implement faster communications protocols and networking equipment, dislocations and associated inefficiencies would likely be more prevalent. 
Opportunity cost realized by market participants (in the form of ROC as detailed above) is ultimately attributable to the physically- and topologically-fragmented nature of the NMS.
Despite this fact, we believe that the current market configuration offers many benefits over alternative configurations, such as the null model defined in Section \ref{sec:lat-arb}.
These results should not be considered as evidence for or against a specific market configuration since, as stated above, the observed phenomena may incentivize the participation of certain kinds of market actors.

We focused our attention on the Dow 30 during calendar year 2016 in order to provide a strong, but tractable baseline.
Future work should investigate longer time periods, larger groups of equities, and other exchange traded products such as Exchange Traded Funds (ETF).
For example, an extension of the current work to larger groups of equities, such as the S\&P 500 or the Russell 3000 would provide greater context for how fragmentation effects different portions of the equities market.
While a time series analysis of dislocation segments and realized opportunity cost series over several years could provide useful information about how fragmentation effects have evolved due to changes in regulation, technology, and market participant behavior.


\section*{Acknowledgements}
The authors gratefully acknowledge helpful discussions with Anshul Anand, Yosry Barsoum, Aaron Blow, Lashon Booker, Robert Books, David Bringle, Richard Byrne, Peter Carrigan, Gary Comparetto, Bryanna Dienta, Peter Sheridan Dodds, Jordan Feidler, Andre Frank, Chase Garwood, Bill Gibson, Frank Hatheway, Emily Hiner, Chuck Howell, Eric Hunsader, Robert Jackson, Neil Johnson, Connie Lewis, Phil Mackintosh, Rishi Narang, Joseph Saluzzi, Wade Shen, Jonathan Smith, Thomas Wilk and all of the participants of the "Flash Mob Research in Computational Finance" hosted by the Complex Systems Center at the University of Vermont.
All opinions and remaining errors are the sole responsibility of the authors and do not reflect the opinions nor perspectives of their affiliated institutions nor that of the funding agencies.
B.F.T., D.R.D, C.M.V.O., J.H.R., and J.V. were supported by DARPA award \#W56KGU-17-C-0010.  B.F.T., M.T.T.K., M.T.M., D.S., and J.V. were supported by HSARPA award \#HSHQDC-14-D-00006. C.M.D.\ was supported by NSF grant \#1447634. C.M.D., and T.J.G.\ were supported by a gift from Massachusetts Mutual Life Insurance Company.
The authors declare no conflicts of interest.
The views, opinions and/or findings expressed are those of the authors and should not be interpreted as representing the official views or policies of the Department of Defense or the U.S. Government.

\bibliographystyle{unsrt}
\bibliography{arxiv_submission}

\appendix

\section{Market participants}
\label{sec:market-participants}
\subsection{Traders}
The four broad classes of traders have different market objectives, and thus generally have different mechanisms for interacting with the market.
Retail investors generally have a small amount of capital and thus interact with the market indirectly, usually through their broker. 
Since their orders are so small in relation to both the value of total market transactions and the size of the inventory of the executing broker-dealer, their orders are often \textit{internalized} (\textit{i.e.} matched against the inventory of their broker-dealer rather than finding a counter-party in the lit market) \cite{sectradeexec}.
For example, if investor A wishes to sell 100 shares of AAPL at the market price to sell, it is likely that, within a large brokerage, investor B wishes to buy 100 shares of AAPL at the market price to buy; these orders can then be executed at the midpoint of the prices.
Alternatively, the broker-dealer may choose to be the counter-party to both traders, using its own inventory of equities and capital.

Institutional investors represent an institution, such as a large corporation, university, or state pension fund.
They are typically far more highly capitalized than a retail investor, and thus their orders are likely to interact more directly (whether through their brokerage or with a market maker) with the NMS.

Brokerages execute orders on behalf of their clients.
They may do this by contacting market-makers, who will execute trades on behalf of the brokerage, or they may themselves be a market-maker or broker-dealer.
Brokerages may enter into contracts with market-makers, who agree to buy some percentage of the brokerage's order flow \cite{sec2007order}.
Securities and Exchange Commission (SEC) (the chief regulatory body of equity markets in the US) regulation requires brokers to guarantee their clients the ``best execution" for their trades, which may include most competitive price for their trades (i.e., highest possible bid price and lowest possible offer price) \cite{secbestexec}.
As has been previously identified elsewhere \cite{nanex_einstein, angel2014finance}, this regulatory requirement for ``Best Price" execution fails to consider implications from special relativity \cite{einstein1905electrodynamics}; namely, that it is impossible to determine whether two distinct events occur at the same time if those events are geographically separated in space.

Market-makers are responsible for ensuring the market's liquidity.
They quote a buy and sell price for a set of traded assets at all times, and stand ready to buy or sell an amount of those assets at their respective prices \cite{sec2000market}.
Exchanges can establish designated market-makers, who are responsible for``making the market'' in a specific asset\cite{menkveld2013high}.
\subsubsection{Market centers}
Exchanges in the NMS are privately-owned venues on which securities are traded.
They are extensively regulated by the SEC and are required by law to provide the best possible execution price (under most circumstances) to their customers \cite{reg-nms, secbestexec}.
For each equity, each exchange maintains a local order book that aggregates the orders submitted by market participants.
These local order books contain information about resting limit orders, updated by order flow, including the side (buy/sell), limit price, size, and execution modifiers that give the market participant greater control over how and when their order is executed.
Using their local book and proprietary matching software, exchanges match buyers with sellers. 

In 2016 there were 13 major stock exchanges:
\vspace{-\topsep}
\begin{itemize}
\item[a)] NYSE (3): main exchange; ARCA, primarily for trading exchange-traded funds (ETFs); and MKT, the smallest of the NYSE family
\item[b)] NASDAQ (3): main exchange; BX, the Boston stock exchange; and PSX, the Philadelphia stock exchange
\item[c)] BATS (4): BATS and BATS Y; EDGX and EDGA. These exchanges are now owned by CBOE, effective early 2017.
\item[d)] IEX: the Investors Exchange, which was a ATS until 17 June 2016
\item[e)] CHX: the Chicago stock exchange
\item[f)] NSX: the National Stock Exchange and by far the smallest stock exchange in terms of shares traded.
It has a long history of trading intermittently, with pauses in operation of duration longer than a year.
\end{itemize}
Though each exchange keeps offices in its namesake city, trading actually occurs (via each exchange's matching engine) in one of three data centers in northern New Jersey; see Section \ref{subsec:physical}. 

ATSs, colloquially known as ``dark pools", are market centers on which invited participants may trade equity and other securities.
While regulated by the SEC, ATSs are not required to publish quotes and are subject to less scrutiny than are the exchanges. 
ATSs are not required to publish the location of their matching engine(s), and as a rule their location is generally not known to the public.
Public SEC filings contain a location for each registered ATS, though it may simply be an office and not the location of the matching engine.
\subsection{Regulatory mechanisms}
The National Market System is regulated primarily by the SEC.
The equities industry also self-regulates through the Financial Industry Regulatory Authority (FINRA), which charges itself with regulating member brokerages and exchanges. 
While an authoritative institution, it does not have law enforcement power itself and must refer suspected violations of securities law to the SEC for enforcement.
(FINRA has some ability to provide incentives and penalties to member organizations, such as expulsion.)
The Securities Information Processor (SIP), mandated by SEC regulation, is a digital information processor on which all quotes, trades, and administrative messages such as trading halts and limit-up / limit-down (LULD) messages are recorded and through which information can be disseminated to exchanges, ATSs, and other market participants.
The SIP constructs the NBBO from this data, which forms the basis of the notion of ``best price'' for the National Market system. 
There are three SIP data collection ``tapes'', two of which (A and B) are located at the NYSE data center in Mahwah, NJ, and one of which (C) is located at the NASDAQ data center in Carteret, NJ.

In addition to the SIP tapes mentioned above, there are two FINRA-operated Trade Reporting Facilities (TRFs), one each in Mahwah and Carteret.
ATSs are required to report trades to the TRFs, which in turn report the trades to the correct SIP tape \cite{sec1998reg-ats, sec2015reg-ats}.

Other regulatory machinery exists to prevent the ``overheating'' of markets in the form of price changes deemed excessive \cite{circuit-breakers}.
There are two types of these mechanisms: individual-stock limit-up, limit-down (LULD) mechanisms and market-wide circuit-breakers. 
Individual stock LULD mechanisms set price bands of 5\%, 10\%, and 20\% for each individual stock based on prices in the immediate trailing five-minute trading period. 
If the stock's price exits the bands and does not return within a fifteen second time period, a five-minute trading halt for that stock is initiated.
Similarly, market-wide circuit breakers (set at 7\%, 13\%, and 20\%) initiate halts in trading if the S\&P 500 breaches these bands. 
A breach of the first two levels results in a market-wide trading halt for 15 minutes, while a breach of the last band results in a trading halt for the rest of the trading day.

Regulatory influence on the market is not limited to price reporting and circuit-breaker mechanisms. 
Beginning in 2016, the SEC instituted a live-market experiment in which some securities would be quoted in minimum increments greater than a penny (which is the current minimum increment at which prices are quoted for all stocks with a share price greater than \$1.00) \cite{tick-pilot}.
Known as the tick-size pilot program (or tick pilot), this program directly alters the pricing mechanism and fundamental price quantization and thus may have an effect on market dynamics.

\section{Glossary and definitions}
\label{sec:definitions}

\subsection*{Market Architecture}
\begin{definition}[\textit{Market System}]\label{def:market_system}
	A market system may be defined as a network or graph which consists of a set of one or more market $centers$ connected by a set of communication channels or ($links$), i.e.\ $system = (centers,\ links)$.
\end{definition}

\begin{definition}[\textit{Market Center}]\label{def:market_center}
	A market center is a location, physical or digital, where agents may interact with a market system.
	A market center may be defined as a tuple containing a local order $book$, a set of valid $actions$, and a set of traded financial $instruments$, i.e.\ $center = (book,\ actions,\ instruments)$.
\end{definition}

\begin{definition}[\textit{Local Order Book}]\label{def:order_book}
	The local order book contains information about the unfulfilled orders that have been submitted to a market center, allowing it to accumulate and maintain state.
	One possible representation of a local order book for a single financial instrument is two ordered lists of queues, where each list is associated with a side of the marked (bid/offer) and each queue is associated with a price.
\end{definition}

\begin{definition}[\textit{Action Set}]
	The action set defines the valid actions at a market center.
	No requirements are imposed on the action set, though a simple real world action set might allow for the submission of limit orders (which guarantee price),  market orders (which guarantee execution), modification of resting orders, and cancellation of resting orders; i.e.\ $actions = \{$limit order, market order, modify, cancel$\}$.
\end{definition}

\begin{definition}[\textit{System Activity}]
	Let the system activity, $\mathbb{A}$, be a chronological list of all actions that are performed in a market system.
	This includes actions performed by market participants, administrative messages transmitted by regulators, and messages transmitted by the exchange(s).
\end{definition}

\begin{definition}[\textit{Data Feed}]\label{def:data_feed}
	A data feed, $D$, is defined to be any subset of the system activity of a market system (i.e.\ $D \subseteq \mathbb{A}$).
	Note that recorded occurrence times of identical events may vary between distinct data feeds due to physical considerations such as the finite speed of information propagation, desynchronized clocks, etc.
\end{definition}

\subsection*{Financial Instruments}
\begin{definition}[\textit{Security}]
	A security is a financial instrument that represents partial or total ownership of an object or entity.
	Securities are fungible; securities belonging to the same ``class'' have the same value, and therefore are interchangeable.
    Additionally, the exact value of a security is negotiable.
	Common varieties of securities include stocks, bonds, and options, all of which may be traded on electronic markets, such as the NMS.
\end{definition}

\begin{definition}[\textit{Stock}]
	Stocks, which are also called equities or equity securities, are a variety of security that represents partial ownership of a publicly traded company.
	Stocks are a vehicle by which companies can acquire the capital necessary to grow and the secondary market for stocks is the basis of a large portion of the U.S. financial industry.
\end{definition}

\subsection*{The Best Bid/Offer}\label{def:lbbo}
The following definitions assume the existence of a market system, $system = (centers,\ links)$.
Each $center \in centers$ has an $action$ set that allows for limit orders and trades a financial instrument $i$.
Additionally, there exists a data feed, $D$, that contains information about the top of the book at each market center (i.e., a consolidated quote feed).
\begin{definition}[\textit{Local Best Bid/Offer}]
    The local best bid and offer (LBBO) is a tuple composed of the local best bid and the local best offer at a particular market center.
	
	The local best bid for $i$ at a particular $center \in centers$, at a time, $t$, is given by the tuple $(p, q)$, where $p$ is the maximum price among all active bids for $i$ in the $book$ at $center$ (as observed via data feed $D$) and $q$ is the quantity of shares of $i$ available at that price at $center$ (i.e.\ $LBB(D, center, i, t) = (p, q)$).
	The local best offer is defined similarly, but uses the minimum price among active offers at $center$ along with the number of shares associated with that order (i.e.\ $LBO(D, center, i, t) = (p', q')$).
\end{definition}

\begin{definition}[\textit{Global Best Bid/Offer}]\label{def:gbbo}
    The global best bid and offer (GBBO) is a tuple composed of the global best bid and the global best offer at a particular market center.
    
	The global best bid is similar to the local best bid, but is formed by the maximum price (and the quantity associated with that order) among resting bids for $i$ among all market $centers$, i.e. $GBB(D, i, t) = (p'', q'')$.	
    Similarly, the global best offer is formed by the minimum price among resting offers and the number of shares at that price (i.e. $GBO(D, i, t) = (p''', q''')$).

    The NBBO, provided by the SIP, is an example of a GBBO in the NMS.
    Note that any real implementation of a GBBO necessitates the introduction of some amount of latency from propagation delays between the market centers and consolidating entity.
    This latency can have material implications in electronic markets where information propagation approaches the speed of light.
\end{definition}

\subsection*{Market Inefficiencies}\label{sec:inefficiencies}
The following definitions assume the existence of a market system, $system = (centers,\ links)$, containing two market centers, two data feeds, $D_1$ and $D_2$, and a financial instrument $i$ that is traded at each $center \in centers$.
$D_1$ and $D_2$ are assumed to contain quote information from each market center, though they may have additional information that contributes to their uniqueness.
Additionally, the distribution of reporting latency and timestamps associated with each event may differ between the feeds. \\
Note that these definitions are phrased for the best bid, but apply similarly to the best offer.

\begin{definition}[\textit{Price Discrepancy}]
	A bid price discrepancy is said to occur when the best bid price differs between $D_1$ and $D_2$, i.e.\ $$\Delta BB(i, t) = BB(D_1, i, t).price - BB(D_2, i, t).price \neq 0.$$
\end{definition}

\begin{definition}[\textit{Market Inefficiency}]
	A market inefficiency occurs whenever a market participant is able to systematically profit from a price discrepancy, usually via the purchase and immediate sale of i.
\end{definition}

\begin{definition}[\textit{Dislocated Data Feeds}]
	$D_1$ and $D_2$ are dislocated with respect to the best bid of $i$ at a time $t$ if there is a bid price discrepancy between $D_1$ and $D_2$.
\end{definition}

\begin{definition}[\textit{Dislocation}]\label{def:dislocation}
	A dislocation between $D_1$ and $D_2$ occurs whenever they are \textit{dislocated} with respect to the best bid of $i$ over a half-open interval of time $[a, b)$.
\end{definition}

\begin{definition}[\textit{Differing trade}]\label{def:differing-trade}
A trade is referred to as a differing trade if it occurs during the lifetime of a dislocation.
\end{definition}

\begin{definition}[\textit{Dislocation Segment}] \label{def:arb_opp}
	A dislocation segment with respect to the best bid of $i$ is any half-open interval of time, $[a, b)$, where $D_1$ and $D_2$ are dislocated with respect to the best bid of $i$ and $\sgn(\Delta BB(i, t)) = \sgn(\Delta BB(i, a)) \ \forall t \in [a, b)$.
\end{definition}

\begin{definition}[\textit{Direction}]
	The direction of a dislocation segment over an interval $[a, b)$ is defined as $\sgn(\Delta BB(i, a))$.
\end{definition}

\begin{definition}[\textit{Duration}]
	The duration of a dislocation or dislocation segment over an interval $[a, b)$ is defined as $b - a$. 
\end{definition}

\begin{definition}[\textit{Magnitude}]
	The magnitude of a dislocation or dislocation segment over an interval $[a, b)$ may be defined as one of the following:
    \begin{align*}
    max\_mag &= \max_{t \in [a, b)}\{|\Delta BB(i, t)|\} \\
    min\_mag &= \min_{t \in [a, b)}\{|\Delta BB(i, t)|\} \\
    mean\_mag &= \dfrac{max\_mag + min\_mag}{2}
    \end{align*}
\end{definition}

\begin{definition}[\textit{Realized Opportunity Cost}]\label{def:roc}
The Realized Opportunity Cost (ROC) experienced by market participants over a period of time $[a, b]$ is defined as:
$$
\sum_{t \in T} |p_{D_1}(\text{time}(t), \text{side}(t)) - p_{D_2}(\text{time}(t), \text{side}(t))|,
$$
where $T$ are all trades that occurred at the NBBO in the period $[a, b]$, $\text{time}(t)$ is a function that returns the time that trade $t$ executed, $\text{side}(\cdot)$ returns the opposite side (bid or offer) of the order that instigated the trade, $p_{D_1}(\text{time}, \text{side})$ returns the best price displayed on feed $D_1$ at the given time and on the given side, 
and 
$p_{D_2}(\text{time},\text{side})$ 
provides the same information for feed $D_2$.
\end{definition}

\subsection*{Market Actions} \label{sec:market actions}
	The following definitions provide a high-level description of the purpose and details of some common order types, but are not necessarily representative of implementations at NMS market centers.

\begin{definition}[\textit{Limit Order}] \label{def:limit order}
	Guarantees market participants an execution price no worse than a provided limit price, but does not provide any guarantees about the timeliness of execution.
	This may be implemented by placing a received limit order into the price queue associated with the provided limit price on the correct side of the book (bid or offer, as specified by the order), assuming that it did not match with a resting order at a better price.\\
	Fields: Instrument identifier, bid/offer, limit price, desired quantity.
\end{definition}

\begin{definition}[\textit{Market order}] \label{def:market order}
	Guarantees instant execution on a best effort basis, but does not provide any guarantees about the execution price.
	This may be implemented by matching the market order with the best resting orders on the opposite side of the book until the desired quantity is obtained.
	A market order may be thought of as a limit order with the limit price set in order to guarantee execution (i.e. 0 for a market offer or infinity for a market bid).\\
	Fields: Instrument identifier, bid/offer, desired quantity
\end{definition}

\begin{definition}[\textit{Modify}]
	Allows market participants to update values associated with resting orders and allows for adaptation to changing market conditions.
	The main usage of this order is to change the number of shares required to fulfill a particular order, since modifying the limit price of order may cause it to lose its place in its current price queue. \\
	Fields: Order identifier, field(s) to modify, new value(s)
\end{definition}

\begin{definition}[\textit{Cancel}]
	Allows market participants to remove resting orders from the local book prior to execution.\\
	Fields: Order identifier
\end{definition}

\begin{definition}[\textit{Immediate Or Cancel}] \label{def:IOC}
	Often shortened to IOC, this is a modifier which may be applied to any order rather than a stand alone order type.
    The modifier indicates that the associated order should be executed immediately upon receipt or canceled if immediate execution is not possible.
\end{definition}

\begin{definition}[\textit{Non-Displayed Orders}]
Orders may be marked with a conditional flag which indicates that they should not be displayed on an exchanges order book, in part or whole.
Such orders are sometimes referred to as hidden orders, since market participants can not identify active non-displayed orders in an order book from publicly available information.

Non-displayed orders may come with some negative consequences including increased fees and decreased execution priority in comparison with displayed orders with identical attributes.
\end{definition}

\begin{definition}[\textit{Midpoint Peg}]
A variety of hidden order that executes at the midpoint of the NBBO, i.e. $0.5(NBB.price + NBO.price)$.
\end{definition}

\section{Regulation National Market System} \label{sec:reg-nms}
Regulation National Market System (Reg.\ NMS) is the set of regulations which defines much of the macro-level organization of the U.S. NMS.
The primary goal of Reg.\ NMS is the creation a unified National Stock Market, additionally it has two secondary goals: to promote competition between markets and between orders, and to serve the interests of long-term investors and listing companies \cite{reg-nms}.
Reg.\ NMS is composed of several rules and regulations, the most important of which are summarized below.
See \cite{reg-nms} for more details.

\subsection*{Order Protection Rule}
The Order Protection Rule (Rule 611), also known as the Trade-through Rule, is meant to protect orders from trade-throughs, which occur when a market center matches an order against a local counter-party when a better price is available via a protected quotation displayed by an alternative market center.
Note that a ``better'' price in this context is defined from the perspective of the new order entering the market, a.k.a. a liquidity demanding or liquidity consuming order.
Therefore a lower execution price is be considered better for an entering bid (offer to buy), while a higher execution price is be considered better for an entering offer (offer to sell).

A protected quotation is defined in Reg. NMS as a bid or offer quotation that satisfies the following properties: the quotation must be automated, the quotation must be displayed by an automated trading center, and the quotation must offer the lowest offer price or highest bid price among all publicly displayed quotations.

A quotation is considered automated if it may be executed without human intervention (up to the full listed quantity), allows for the correct execution of Immediate-Or-Cancel (IOC) orders against the quotation, immediately provides a response to the sender of an Immediate-Or-Cancel order indicating the execution status of that order, and immediately updates the quotation to reflect any changes to its status.

A trading center is considered automated if it implements systems and procedures that allow it to display automated quotations as defined above, and quotations that do not satisfy the requirements of an automated quotation are identified as manual quotations as quickly as possible.

Trade-throughs are prohibited Under Rule 611, however exceptions are allowed for Inter-market Sweep Orders (ISO), quotations displayed by markets that fail to meet the reporting requirements for automated quotations, and flickering quotations with multiple prices displayed in a single second.

\subsection*{Access Rule}
The Access Rule (Rule 610) concerns itself with setting standards for access to quotations in NMS stocks, and caps the fees that an exchange may charge for accessing its protected quotations at $\$0.003$ per share.
Rule 610 allows for the creation and usage of private data feeds, often referred to as direct feeds by market participants since they are offered directly by exchanges rather than through a third party.
Rule 610 also prohibits trading centers from displaying quotations which would lock or cross a protected quotation from a different trading center.

A market is said to be locked if the bid-offer spread of that market is zero, in other words there exists a resting bid and a resting offer with identical limit prices.
A market is said to be crossed if the bid-offer spread of that market is negative, i.e. there exists a resting bid whose limit price is greater than the limit price of a resting offer, or equivalently a resting offer exists whose limit price is less than the limit price of a resting bid.
These effects are the result of coupling geographically fragmented exchanges, since an order that may lock or cross a market would immediately find a counter-party if the two orders were present on the same exchange.

\subsection*{Sub-Penny Rule}
The Sub-Penny Rule (Rule 612) prohibits market participants from displaying or accepting quotations for NMS stocks priced in an increment less than $\$0.01$ unless the quotation price is less than $\$1.00$, in which case the minimum increment is $\$0.0001$.
Rule 612 is meant to prohibit the practice of ``sub-pennying'' in which market participants could ``step ahead'' of a protected quotation by providing a negligible amount of price improvement, allowing the ``sub-pennied'' order faster execution at effectively no extra cost.

The significance of this rule, with respect to geographic fragmentation and market inefficiencies, is that the minimum increment for the quoted price of a traded instrument sets the minimum magnitude of all
dislocation segments.

\subsection*{Market Data Rules}
Rules 601 and 603 are referred to as Market Data Rules and are meant to promote wide availability of market data, thus providing all market participants with an accurate and reliable source of information on the best prices in NMS stocks.
These rules cover the organization of a consolidated data feed for NMS stocks, the reward structure for contributing information to the consolidated data feed, and establishes standards for quote and trade information provided to and provided by the consolidated data feed.

In particular these rules concern the Consolidated Tape Association (CTA) plan which disseminates transaction information for NYSE listed securities, the Consolidated Quotation (CQ) plan which disseminates quote information for NYSE listed securities, and the Nasdaq UTP plan which disseminates quote and trade data for Nasdaq listed securities.
The information provided by the CTA plan and CQ plan forms Consolidated Tape A, and the information provided by the UTP plan forms Consolidated Tape C.
There also exists a Consolidated Tape B which reports trade information for stocks listed on regional exchanges.
The aggregation of Consolidated Tapes A, B, and C form what is commonly referred to as the SIP feed.

\clearpage
\section{Figures and Tables}

\footnotesize
\begin{longtable*}[c]{ | llrrrrrr | }
\hline
Symbol &       &  Trades &    Traded Val &  Diff Trades &  Diff Traded Val &       ROC &  ROC/Share\rule{0pt}{2.6ex} \\
\hline
\endhead
\hline
\multicolumn{8}{|r|}{{Continued on next page\rule{0pt}{2.6ex}}} \\ \hline
\endfoot

\endlastfoot
AAPL & mean & 174,820.85 & 2,542,188,952.00 &     34,316.58 &      483,265,898.89 &    45,852.81 &   0.007569\rule{0pt}{2.6ex} \\
& std &  68,897.09 & 1,040,923,482.82 &     20,556.77 &      280,321,422.73 &    27,275.35 &   0.001224 \\
& min &     54,824 &   983,856,430.54 &         2,112 &       35,009,317.38 &     2,773.35 &   0.004007 \\
& 25\% &    129,830 & 1,872,512,861.35 &     23,498.75 &      340,459,129.85 &    32,088.96 &   0.007307 \\
& 50\% & 156,198.50 & 2,272,037,106.11 &        32,741 &      452,246,993.35 &    43,204.10 &   0.007682 \\
& 75\% & 199,793.25 & 2,870,019,105.28 &     42,674.75 &      599,146,631.96 &    57,647.05 &   0.008351 \\
& max &    517,270 & 8,280,915,338.59 &       103,885 &    1,596,912,962.05 &   138,331.08 &   0.011924 \\
AXP & mean &  32,348.46 &   250,614,304.97 &      9,086.69 &       71,464,081.61 &    11,622.14 &   0.008811 \\
& std &  16,110.77 &   143,031,721.64 &      4,434.64 &       36,283,858.01 &     7,156.73 &   0.000757 \\
& min &     11,095 &    90,438,986.65 &         2,219 &       19,241,382.52 &     2,666.91 &   0.007285 \\
& 25\% &  22,756.50 &   168,209,590.34 &      5,999.50 &       49,149,197.52 &     7,672.38 &   0.008324 \\
& 50\% &     26,835 &   207,178,850.49 &         7,476 &       57,481,058.84 &     8,987.37 &   0.008723 \\
& 75\% &  37,067.75 &   277,456,051.09 &     10,905.75 &       86,485,488.61 &    13,792.36 &   0.009248 \\
& max &    159,135 & 1,468,245,304.80 &        31,507 &      302,294,385.78 &    75,473.73 &   0.013393 \\
BA & mean &  20,749.26 &   288,851,358.58 &      7,071.25 &      100,312,506.47 &    10,955.15 &   0.012812 \\
& std &  10,435.29 &   154,859,396.19 &      3,027.87 &       41,626,820.43 &     6,235.13 &   0.008003 \\
& min &      4,220 &    60,869,511.81 &         1,209 &       22,712,059.97 &     2,404.84 &   0.008497 \\
& 25\% &  14,825.75 &   202,629,761.69 &         4,865 &       69,859,447.55 &     6,607.66 &   0.010588 \\
& 50\% &     18,904 &   260,864,798.97 &      6,613.50 &       95,165,851.81 &     9,608.32 &   0.011730 \\
& 75\% &  24,641.25 &   339,733,518.51 &      8,877.75 &      123,131,081.82 &    13,061.48 &   0.013252 \\
& max &    101,159 & 1,496,951,020.26 &        19,630 &      303,000,376.46 &    47,010.92 &   0.131181 \\
CAT & mean &  30,586.73 &   269,579,023.84 &      9,239.74 &       81,143,988.26 &    11,986.17 &   0.010142 \\
& std &  11,384.23 &   107,296,519.47 &      3,721.88 &       30,953,831.34 &     8,988.83 &   0.005617 \\
& min &      7,660 &    72,342,016.91 &         2,283 &       24,025,499.19 &     2,847.50 &   0.007044 \\
& 25\% &  22,670.50 &   204,956,948.21 &      6,684.75 &       58,633,048.01 &     7,680.36 &   0.008842 \\
& 50\% &     28,267 &   245,802,664.58 &         8,451 &       76,013,433.31 &    10,301.82 &   0.009394 \\
& 75\% &  36,304.25 &   323,347,949.42 &        10,730 &       95,123,308.69 &    13,455.58 &   0.010105 \\
& max &     77,886 &   964,799,514.35 &        22,381 &      222,261,612.89 &   100,244.92 &   0.084153 \\
CSCO & mean &  77,364.30 &   493,693,519.98 &     11,555.12 &       74,134,548.33 &    26,409.30 &   0.008899 \\
& std &  33,235.82 &   207,062,395.07 &      8,173.36 &       50,695,319.75 &    19,401.61 &   0.001000 \\
& min &     31,865 &   182,535,557.30 &           660 &        4,502,758.25 &     1,461.77 &   0.005896 \\
& 25\% &     58,015 &   367,489,467.23 &         6,881 &       46,381,850.87 &    15,394.81 &   0.008321 \\
& 50\% &  68,328.50 &   444,190,912.86 &        10,643 &       70,264,638.23 &    23,922.43 &   0.009015 \\
& 75\% &  86,368.50 &   548,980,902.84 &     14,364.50 &       92,558,544.11 &    32,439.85 &   0.009212 \\
& max &    307,808 & 1,702,786,754.09 &        58,922 &      316,907,129.91 &   130,317.79 &   0.019144 \\
CVX & mean &  44,441.79 &   462,460,384.39 &     12,439.81 &      134,648,014.78 &    17,036.07 &   0.012786 \\
& std &  17,816.08 &   164,739,606.44 &      6,693.09 &       56,874,558.81 &    16,228.50 &   0.034249 \\
& min &     13,879 &   144,582,207.22 &         2,377 &       28,830,654.89 &     2,456.15 &   0.006135 \\
& 25\% &  32,594.50 &   346,722,417.91 &      8,344.75 &       97,784,127.50 &     9,772.81 &   0.008240 \\
& 50\% &  39,655.50 &   430,819,298.93 &     10,794.50 &      122,659,276.18 &    12,993.21 &   0.008796 \\
& 75\% &  53,123.50 &   538,846,282.98 &     14,257.25 &      158,567,573.75 &    18,798.64 &   0.009602 \\
& max &    148,515 & 1,263,782,534.87 &        50,186 &      423,871,063.95 &   190,901.32 &   0.531465 \\
DD & mean &  18,036.06 &   132,521,012.09 &      4,913.74 &       37,476,052.09 &     6,342.15 &   0.009882 \\
& std &   8,759.67 &    63,295,360.33 &      2,764.13 &       19,045,796.05 &     4,403.69 &   0.004642 \\
& min &      5,262 &    40,582,912.43 &           773 &        6,491,584.69 &     832.8600 &   0.004446 \\
& 25\% &     12,017 &    89,557,470.32 &      3,123.50 &       24,611,595.27 &     3,912.60 &   0.008929 \\
& 50\% &     15,462 &   114,690,819.55 &      4,104.50 &       32,687,710.05 &     5,066.52 &   0.009511 \\
& 75\% &  20,793.50 &   155,332,165.60 &         5,499 &       44,016,133.46 &     7,243.21 &   0.010036 \\
& max &     52,298 &   418,605,566.86 &        15,217 &      113,435,890.12 &    42,392.91 &   0.080300 \\
DIS & mean &  41,156.78 &   495,392,306.10 &     10,535.97 &      129,495,234.75 &    39,331.64 &   0.024431 \\
& std &  15,686.14 &   208,901,188.83 &      4,550.24 &       53,409,308.00 &   323,220.91 &   0.153256 \\
& min &     17,030 &   203,854,389.74 &         2,633 &       36,156,641.31 &     3,221.94 &   0.006826 \\
& 25\% &  31,892.25 &   374,803,933.38 &      7,657.50 &       97,517,442.34 &    10,284.53 &   0.008200 \\
& 50\% &  36,745.50 &   430,039,189.49 &      9,220.50 &      114,691,273.09 &    13,055.17 &   0.008814 \\
& 75\% &  45,623.25 &   558,118,900.42 &        11,989 &      144,599,501.14 &    17,818.25 &   0.010073 \\
& max &    124,145 & 1,659,028,038.95 &        32,212 &      369,007,239.36 & 5,138,897.26 &     2.4261 \\
GE & mean &  83,963.26 &   741,830,493.33 &     12,828.05 &      119,789,470.34 &    44,606.60 &   0.011460 \\
& std &  35,661.52 &   309,010,418.40 &      8,012.21 &       69,234,760.11 &    71,027.27 &   0.009108 \\
& min &     27,905 &   290,466,991.56 &         2,653 &       33,035,264.68 &     7,844.38 &   0.005032 \\
& 25\% &  59,365.25 &   514,268,974.25 &         7,603 &       74,660,777.81 &    23,089.88 &   0.008105 \\
& 50\% &  74,767.50 &   670,261,948.27 &        10,589 &       96,944,717.80 &    29,655.10 &   0.009088 \\
& 75\% &     96,517 &   876,527,780.45 &        14,826 &      141,143,821.44 &    45,447.71 &   0.009994 \\
& max &    236,395 & 1,961,985,442.37 &        49,675 &      427,596,291.36 & 1,020,533.87 &   0.074863 \\
GS & mean &  16,072.52 &   266,630,735.82 &      6,039.60 &      100,455,871.70 &    12,632.51 &   0.018917 \\
& std &   6,759.46 &   124,491,943.62 &      2,299.09 &       38,813,844.47 &     7,817.49 &   0.008519 \\
& min &      5,914 &   106,821,197.38 &         1,908 &       43,864,040.44 &     4,126.93 &   0.009892 \\
& 25\% &  11,672.50 &   178,117,450.68 &      4,400.50 &       72,797,279.72 &     8,094.88 &   0.015293 \\
& 50\% &     14,285 &   224,601,809.66 &      5,593.50 &       89,806,560.53 &    10,478.40 &   0.017610 \\
& 75\% &  18,995.50 &   329,800,789.60 &      7,207.25 &      123,468,835.06 &    14,081.00 &   0.020349 \\
& max &     50,816 &   857,877,495.97 &        14,393 &      247,177,637.53 &    72,612.29 &   0.117195 \\
HD & mean &  27,728.62 &   366,840,862.69 &      8,920.89 &      123,442,984.04 &    12,744.17 &   0.011027 \\
& std &   8,963.39 &   127,799,636.38 &      3,551.59 &       45,816,593.86 &    13,953.86 &   0.005344 \\
& min &     13,006 &   165,434,810.90 &         2,515 &       36,439,575.01 &     2,864.63 &   0.007334 \\
& 25\% &  21,668.50 &   276,025,739.12 &         6,473 &       92,772,210.99 &     7,812.18 &   0.009211 \\
& 50\% &     25,747 &   339,935,686.44 &         8,234 &      115,952,305.04 &     9,863.20 &   0.009837 \\
& 75\% &  31,341.50 &   416,697,720.23 &        10,762 &      146,870,527.07 &    13,201.84 &   0.010717 \\
& max &     64,114 & 1,031,531,952.92 &        22,597 &      291,592,154.20 &   186,403.78 &   0.059372 \\
IBM & mean &  19,503.60 &   283,053,487.10 &      6,540.58 &       97,629,157.53 &    10,322.91 &   0.023045 \\
& std &   7,762.60 &   121,204,978.80 &      3,031.09 &       41,935,403.95 &    11,852.61 &   0.155151 \\
& min &      6,168 &    83,951,134.35 &         1,493 &          24,252,638 &     2,042.64 &   0.007853\rule{0pt}{2.6ex} \\
& 25\% &     14,595 &   209,597,644.74 &         4,586 &       71,732,705.50 &     5,972.78 &   0.010419 \\
& 50\% &     17,729 &   252,167,134.97 &      5,852.50 &       89,826,517.09 &     7,844.79 &   0.011260 \\
& 75\% &  22,431.25 &   328,204,236.66 &      7,532.75 &      111,719,286.16 &    10,385.29 &   0.012460 \\
& max &     59,625 &   972,131,459.03 &        21,810 &      299,050,973.50 &   111,628.46 &     2.4712 \\
INTC & mean &  88,012.92 &   539,061,461.61 &     13,623.27 &       80,485,200.80 &    24,652.76 &   0.008581 \\
& std &  32,133.18 &   218,280,102.40 &      8,604.73 &       50,349,950.02 &    16,048.53 &   0.001366 \\
& min &     25,392 &   174,808,926.57 &           668 &        3,512,129.76 &     906.3800 &   0.003979 \\
& 25\% &  66,319.50 &   409,452,090.75 &      8,564.50 &       53,076,046.07 &    15,370.02 &   0.008123 \\
& 50\% &     81,767 &   493,100,646.52 &        13,526 &       79,046,604.94 &    23,962.08 &   0.008955 \\
& 75\% & 100,219.25 &   601,791,580.25 &     17,608.50 &      104,796,930.94 &    32,243.13 &   0.009165 \\
& max &    233,578 & 1,765,833,707.79 &        48,079 &      318,483,188.44 &    91,380.43 &   0.017641 \\
JNJ & mean &  41,248.16 &   516,784,968.61 &     10,117.01 &      132,739,127.27 &    15,971.14 &   0.011066 \\
& std &  13,010.19 &   163,195,302.28 &      4,751.53 &       54,033,725.20 &    24,562.10 &   0.009799 \\
& min &     15,606 &   194,794,413.45 &         2,156 &       34,113,674.01 &     3,046.94 &   0.006887 \\
& 25\% &  32,847.50 &   413,846,348.61 &      7,231.25 &       98,042,130.74 &     8,347.87 &   0.008033 \\
& 50\% &  38,411.50 &   483,292,741.16 &         8,718 &      117,582,458.80 &    10,975.76 &   0.008545 \\
& 75\% &  45,961.50 &   586,813,347.06 &        11,288 &      153,593,921.79 &    16,623.58 &   0.009356 \\
& max &     94,603 & 1,244,615,527.23 &        32,165 &      338,562,051.69 &   362,771.34 &   0.091514 \\
JPM & mean &  88,003.57 &   801,423,694.85 &     21,356.75 &      193,852,644.59 &    29,550.37 &   0.008671 \\
& std &  39,466.22 &   360,958,601.04 &     11,483.72 &       91,730,373.77 &    14,749.77 &   0.001427 \\
& min &     30,040 &   331,806,293.97 &         4,953 &       58,788,624.46 &     7,089.25 &   0.006291 \\
& 25\% &  61,325.75 &   565,821,050.04 &        13,638 &      130,610,914.09 &    19,065.50 &   0.007994 \\
& 50\% &     77,139 &   711,684,130.50 &     17,913.50 &      171,373,698.77 &    25,663.01 &   0.008471 \\
& 75\% & 101,690.75 &   948,789,239.22 &        25,153 &      232,200,018.53 &    34,390.20 &   0.008981 \\
& max &    256,973 & 3,004,137,079.38 &        70,052 &      646,651,792.53 &    92,386.71 &   0.019550 \\
KO & mean &  52,120.74 &   406,264,869.51 &     10,086.25 &       81,371,474.40 &    18,263.90 &   0.009458 \\
& std &  19,287.46 &   161,269,975.11 &      4,577.72 &       33,628,799.23 &     8,429.36 &   0.003791 \\
& min &     19,958 &   185,384,176.07 &         3,156 &       30,732,830.23 &     7,111.74 &   0.006435 \\
& 25\% &  39,138.50 &   301,353,437.57 &      7,209.25 &       59,076,462.53 &    13,153.95 &   0.008509 \\
& 50\% &  47,536.50 &   368,857,020.57 &         8,995 &       74,612,460.50 &    16,482.05 &   0.008996 \\
& 75\% &     58,796 &   463,324,316.41 &     11,326.50 &       92,283,870.91 &    20,688.83 &   0.009567 \\
& max &    151,901 & 1,308,364,552.46 &        30,895 &      222,649,014.88 &    88,890.33 &   0.059954 \\
MCD & mean &  28,809.30 &   380,847,318.26 &      7,442.77 &      103,499,997.57 &    10,822.55 &   0.010045 \\
& std &   9,250.20 &   146,529,362.71 &      2,681.91 &       39,288,571.37 &    10,847.35 &   0.004427 \\
& min &      9,911 &      117,553,924 &         2,479 &       32,522,381.94 &     2,926.51 &   0.007534 \\
& 25\% &  22,526.25 &   277,305,454.74 &      5,422.50 &       75,412,153.02 &     6,484.23 &   0.008638 \\
& 50\% &  26,999.50 &   355,968,666.10 &      7,088.50 &       98,050,825.62 &     8,795.32 &   0.009242 \\
& 75\% &  33,173.25 &   455,898,847.39 &      8,601.50 &      121,862,623.83 &    11,289.40 &   0.009876 \\
& max &     72,028 & 1,044,773,633.09 &        20,018 &      265,940,261.14 &   114,279.57 &   0.055288 \\
MMM & mean &  11,365.37 &   167,307,657.17 &      3,636.52 &       57,734,183.69 &    12,063.44 &   0.017206 \\
& std &   3,901.98 &    56,357,516.03 &      1,769.07 &       23,315,655.68 &   102,963.80 &   0.055266 \\
& min &      3,704 &    42,376,029.54 &           852 &       12,737,335.17 &     1,268.98 &   0.008572 \\
& 25\% &   8,870.50 &   128,614,072.44 &         2,564 &       42,198,399.30 &     3,302.92 &   0.011656 \\
& 50\% &     10,484 &   156,620,977.62 &         3,148 &       53,116,097.50 &     4,412.75 &   0.012771 \\
& 75\% &     13,011 &   192,588,435.88 &         4,113 &       67,265,920.46 &     6,182.75 &   0.014374 \\
& max &     27,168 &   374,180,512.59 &        11,339 &      141,420,561.60 & 1,638,916.42 &   0.888354 \\
MRK & mean &  52,065.45 &   404,241,094.10 &     12,269.51 &       97,773,420.29 &    17,435.31 &   0.008974 \\
& std &  21,247.82 &   198,964,179.96 &      6,450.80 &       49,864,763.12 &    10,578.05 &   0.002336 \\
& min &     18,727 &   139,953,296.94 &         4,541 &       37,156,365.39 &     5,935.82 &   0.005243 \\
& 25\% &  39,157.50 &   302,275,577.81 &         7,789 &       64,970,853.90 &    10,991.95 &   0.008151 \\
& 50\% &  46,619.50 &   360,181,487.45 &        10,518 &       84,618,791.63 &    15,107.30 &   0.008574 \\
& 75\% &  58,293.50 &   460,791,595.64 &     13,950.25 &      113,701,288.72 &    19,811.20 &   0.008998 \\
& max &    232,717 & 2,584,131,245.57 &        46,595 &      456,348,016.09 &   112,089.37 &   0.027925 \\
MSFT & mean & 141,856.07 & 1,190,901,402.50 &     24,761.04 &      203,129,267.74 &    36,706.48 &   0.008303 \\
& std &  63,588.22 &   533,057,860.34 &     17,480.90 &      139,593,661.04 &    25,629.62 &   0.001006 \\
& min &     37,036 &   459,917,664.02 &         1,070 &        8,596,209.02 &     1,253.33 &   0.004649 \\
& 25\% & 102,602.75 &   837,915,412.53 &     14,050.50 &      122,690,779.06 &    22,117.84 &   0.007944 \\
& 50\% & 124,327.50 & 1,053,837,918.53 &        22,263 &      180,976,495.62 &    32,474.83 &   0.008414 \\
& 75\% &    156,482 & 1,373,674,878.44 &     32,070.75 &      262,037,990.81 &    48,649.14 &   0.008943 \\
& max &    456,106 &    4,125,126,448 &        98,307 &      950,946,403.87 &   138,913.71 &   0.010702 \\
NKE & mean &  46,386.10 &   377,535,172.78 &     10,935.36 &       89,164,054.37 &    18,227.11 &   0.009535 \\
& std &  15,357.30 &   145,806,631.19 &      3,796.12 &       32,750,534.77 &    20,652.07 &   0.006259 \\
& min &     13,818 &    84,721,641.01 &         2,885 &       18,676,669.13 &     3,523.46 &   0.005848 \\
& 25\% &  37,737.50 &   295,226,871.17 &      8,613.25 &       69,144,244.49 &    12,031.69 &   0.008070 \\
& 50\% &     42,544 &   344,601,219.52 &         9,822 &       80,592,736.50 &    14,390.93 &   0.008480 \\
& 75\% &  51,532.50 &   424,862,753.77 &        12,534 &      102,100,476.22 &    18,212.90 &   0.008969 \\
& max &    121,962 & 1,195,681,284.35 &        28,410 &      232,923,873.16 &   280,266.40 &   0.084753 \\
PFE & mean &  91,040.68 &   692,324,391.87 &     13,862.73 &      110,715,986.10 &    31,625.70 &   0.009084 \\
& std &  49,256.08 &   473,362,104.74 &      6,672.49 &       60,406,629.99 &    16,222.08 &   0.002189 \\
& min &     32,599 &   212,898,806.65 &         4,422 &       28,855,501.39 &     8,447.34 &   0.005328 \\
& 25\% &  59,097.50 &   426,001,630.46 &      9,726.75 &       74,658,424.30 &    21,093.39 &   0.008060 \\
& 50\% &     80,628 &   611,356,656.02 &     13,270.50 &      103,824,254.59 &    29,745.09 &   0.008872 \\
& 75\% & 109,044.50 &   783,454,707.50 &        16,379 &      129,458,335.77 &    36,810.46 &   0.009218 \\
& max &    474,221 & 5,427,524,575.47 &        56,238 &      602,885,333.66 &   145,936.99 &   0.021475 \\
PG & mean &  50,438.27 &   570,844,223.65 &     11,760.85 &      134,139,256.91 &    17,786.87 &   0.011319 \\
& std &  26,464.80 &   419,733,122.26 &      5,828.25 &       70,501,433.31 &    13,441.76 &   0.027016 \\
& min &     19,980 &   185,431,171.67 &         3,696 &       40,926,831.50 &     4,789.61 &   0.005871 \\
& 25\% &  34,682.50 &      362,299,048 &      7,530.75 &       87,831,864.68 &    10,158.57 &   0.007830 \\
& 50\% &  43,215.50 &   456,219,304.50 &     10,168.50 &      113,664,173.39 &    14,134.84 &   0.008337\rule{0pt}{2.6ex} \\
& 75\% &     57,796 &   612,257,033.06 &        14,163 &      160,102,759.35 &    20,335.25 &   0.008869 \\
& max &    181,697 & 3,330,428,860.98 &        38,467 &      460,594,145.16 &   111,040.42 &   0.427532 \\
TRV & mean &  10,544.19 &   106,389,400.10 &      3,568.88 &       39,506,286.77 &     4,441.92 &   0.011206 \\
& std &   3,416.58 &    36,241,051.32 &      1,447.62 &       15,393,415.99 &     2,794.22 &   0.002964 \\
& min &      3,018 &    27,592,851.46 &           771 &        7,628,101.68 &     964.9800 &   0.007730 \\
& 25\% &   8,487.25 &    82,492,360.44 &         2,705 &       29,702,903.75 &     2,990.61 &   0.009813 \\
& 50\% &   9,965.50 &   101,071,670.21 &      3,334.50 &       37,475,398.81 &     3,837.38 &   0.010699 \\
& 75\% &  12,010.25 &   122,831,584.80 &      4,172.25 &       46,650,233.97 &     4,933.57 &   0.011895 \\
& max &     27,468 &   294,476,802.95 &        11,339 &      107,591,813.81 &    28,594.17 &   0.048296 \\
UNH & mean &  17,446.67 &   228,660,097.56 &      5,642.65 &       77,377,042.02 &     7,680.73 &   0.011369 \\
& std &   5,246.70 &    81,435,935.28 &      2,011.09 &       27,032,487.15 &     4,216.99 &   0.001956 \\
& min &      6,412 &    89,234,548.68 &         1,849 &       26,225,274.13 &     2,378.89 &   0.008077 \\
& 25\% &     14,129 &   173,512,633.11 &      4,413.50 &       59,089,553.72 &     5,357.03 &   0.010139 \\
& 50\% &     16,636 &   214,637,619.41 &      5,371.50 &       75,046,042.50 &     6,539.35 &   0.010909 \\
& 75\% &  19,932.50 &   260,900,529.94 &      6,717.75 &       92,546,473.56 &     8,912.00 &   0.012157 \\
& max &     41,842 &   725,532,688.10 &        15,652 &      218,550,591.96 &    30,826.07 &   0.020873 \\
UTX & mean &  24,903.26 &   263,375,122.23 &      8,217.23 &       88,158,366.16 &    17,510.92 &   0.011823 \\
& std &  12,739.37 &   141,586,417.29 &      4,913.50 &       47,017,041.73 &   109,897.32 &   0.025491 \\
& min &      5,358 &    49,595,310.95 &         1,315 &       13,549,323.03 &     1,579.70 &   0.007425 \\
& 25\% &     16,977 &   182,655,118.69 &      4,942.75 &       59,862,917.81 &     6,260.83 &   0.009197 \\
& 50\% &     21,463 &   229,710,235.45 &      7,034.50 &       77,766,133.91 &     8,629.70 &   0.009769 \\
& 75\% &  27,806.50 &   295,642,614.43 &      9,447.25 &      101,084,369.56 &    11,780.77 &   0.010558 \\
& max &     86,284 & 1,144,629,181.20 &        29,297 &      275,444,139.17 & 1,749,683.12 &   0.413092 \\
V & mean &  48,950.33 &   460,497,961.48 &     13,097.62 &      122,925,302.52 &    16,818.68 &   0.009524 \\
& std &  17,793.66 &   170,963,876.45 &      6,162.95 &       50,951,708.77 &     9,603.52 &   0.007991 \\
& min &     23,142 &   162,781,451.21 &         3,273 &       30,311,313.92 &     3,873.41 &   0.005686 \\
& 25\% &     36,797 &   351,926,962.18 &      9,092.75 &       88,092,316.21 &    11,157.63 &   0.007977 \\
& 50\% &     44,660 &   411,627,578.41 &        11,552 &      112,507,821.59 &    14,624.21 &   0.008478 \\
& 75\% &     56,347 &   531,962,904.82 &        14,735 &      139,117,677.04 &    18,457.99 &   0.009089 \\
& max &    128,775 & 1,261,830,529.49 &        42,661 &      355,125,487.75 &    85,584.79 &   0.120466 \\
VZ & mean &  62,098.01 &   494,149,523.80 &     13,525.58 &      109,544,287.76 &    51,450.08 &   0.013465 \\
& std &  23,339.73 &   185,520,474.01 &      5,963.55 &       44,308,281.09 &   427,124.11 &   0.030589 \\
& min &     29,671 &   204,408,079.42 &         5,039 &       41,836,595.67 &     6,539.31 &   0.005940 \\
& 25\% &  46,137.50 &   362,469,762.67 &      9,445.50 &       77,277,337.22 &    14,070.98 &   0.008312 \\
& 50\% &  55,823.50 &   449,943,857.51 &        11,922 &      100,842,957.95 &    19,546.38 &   0.008925 \\
& 75\% &     71,345 &   574,383,307.99 &     15,340.50 &      128,753,322.82 &    27,179.16 &   0.009833 \\
& max &    147,919 & 1,264,130,771.03 &        36,340 &      266,067,716.19 & 6,798,041.07 &   0.469146 \\
WMT & mean &  49,823.30 &   448,218,124.11 &     11,786.63 &      109,524,107.26 &    19,815.12 &   0.011010 \\
& std &  20,042.63 &   187,614,765.33 &      5,642.63 &       49,138,231.17 &    26,412.77 &   0.013753 \\
& min &     20,706 &   211,540,076.99 &         3,709 &       34,219,678.86 &     4,605.33 &   0.006303 \\
& 25\% &  36,156.25 &   325,522,820.91 &         7,935 &       74,826,489.33 &    10,770.16 &   0.008199 \\
& 50\% &  44,622.50 &   399,048,171.16 &        10,657 &       99,148,394.58 &    14,630.73 &   0.008728 \\
& 75\% &  57,546.50 &   520,989,325.68 &     13,105.25 &      125,786,171.91 &    19,936.97 &   0.009294 \\
& max &    156,021 & 1,562,166,750.41 &        36,698 &      361,429,655.92 &   246,675.56 &   0.176158 \\
XOM & mean &  64,074.02 &   670,862,447.91 &     17,774.64 &      188,657,442.78 &    35,104.83 &   0.013337 \\
& std &  28,483.97 &   265,569,760.30 &     10,924.37 &       93,450,720.08 &   127,162.28 &   0.021539 \\
& min &     21,646 &   201,555,090.63 &         4,205 &       46,296,094.35 &     4,953.61 &   0.005362 \\
& 25\% &     46,888 &   496,895,704.13 &     11,816.25 &      129,373,837.81 &    15,072.46 &   0.007813 \\
& 50\% &  55,080.50 &   593,690,988.09 &        14,020 &      162,976,539.44 &    18,862.46 &   0.008369 \\
& 75\% &  74,045.75 &   786,147,285.48 &     19,397.50 &      211,792,668.60 &    31,372.43 &   0.010116 \\
& max &    209,816 & 1,761,362,028.61 &        75,421 &      613,405,517.24 & 2,003,841.58 &   0.238129 \\
\hline
\noalign{\vskip 1mm}
\caption{
\normalsize
Summary ROC statistics for Dow 30 stocks, aggregated by day and trading symbol.
}
\label{tab:roc-stats}
\end{longtable*}
\normalsize

\begin{figure*}[tp!]
	\centering	
	\includegraphics[width=\textwidth]{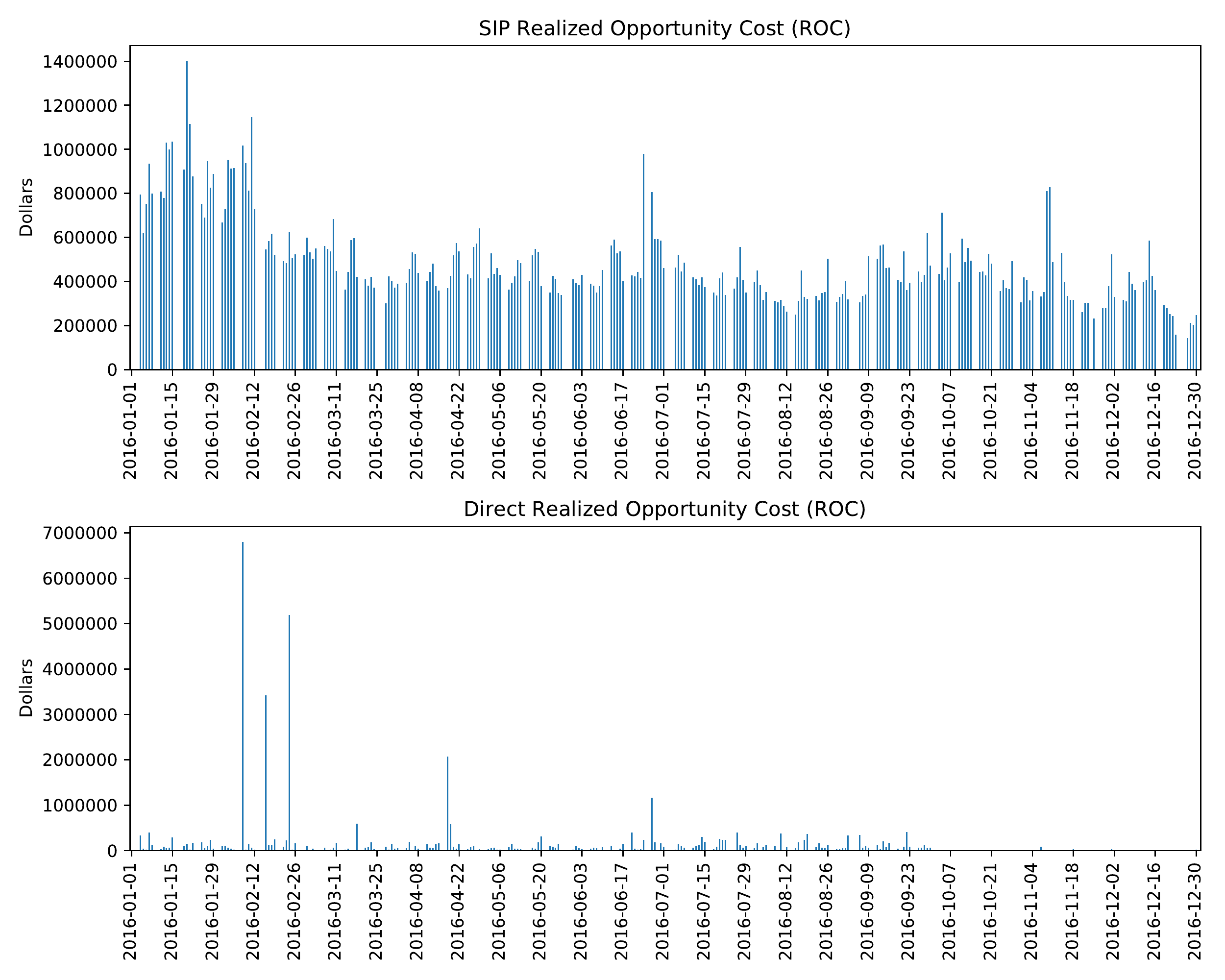} 
	\caption{
		Daily ROC during calendar year 2016 aggregated across all tickers.
		A large majority of days favored the direct data feeds.
		Both Direct and SIP ROC time series show signs of decay across 2016, which may be due to infrastructure improvements.
	}
	\label{fig:purse_by_day}
\end{figure*}

\begin{table*}[!h]
\centering
\textbf{Direct Feed and Historical Data Pricing} \\
\begin{tabular}{| llrr |}
\hline
Data Provider               & Product             & One-time Cost & Monthly Fee \\ \hline
CTA                         & CQS/CTS Feed        &               &   \$1,850   \\
UTP                         & UQDF/UTDF Feed      &               &  *\$2,500   \\
NYSE                        & Integrated Feed     &               &   \$7,500   \\
                            & Historical DoB      & \$60,000      &             \\
                            & Historical ToB      & \$36,000      &             \\
NYSE ARCA                   & Integrated Feed     &               &   \$3,000   \\
                            & Historical DoB      & \$36,000      &             \\
NYSE MKT                    & Integrated Feed     &               &   \$2,500   \\
(Now NYSE American)         & Historical DoB      & \$18,000      &             \\
National Stock Exchange NSX & Integrated Feed     &               &       \$0   \\
(Now NYSE National)         & Historical DoB      &               & **\$1,500   \\
NASDAQ                      & TotalView-ITCH Feed &               & *\$25,000   \\
                            & Historical DoB      &               &  *\$1,250   \\
NASDAQ BX                   & TotalView-ITCH Feed &               & *\$20,000   \\
                            & Historical DoB      &               &    *\$500   \\
NASDAQ PSX                  & TotalView-ITCH Feed &               & *\$17,000   \\
                            & Historical DoB      &               &    *\$500   \\
BATS BZX                    & Depth Feed          &               &   \$2,000   \\
                            & Historical DoB      & \$8,500       &             \\
BATS BYX                    & Depth Feed          &               &   \$2,000   \\
                            & Historical DoB      & \$8,500       &             \\
Direct Edge EDGA            & Depth Feed          &               &   \$1,000   \\
                            & Historical DoB      & \$8,500       &             \\
Direct Edge EDGX            & Depth Feed          &               &   \$2,000   \\
                            & Historical DoB      & \$8,500       &             \\
The Investors Exchange      & TOPS Feed           & \$0           &       \$0   \\
                            & DEEP Feed           & \$0           &       \$0   \\
                            & Historical DoB      & \$0           &       \$0   \\
Chicago Stock Exchange      & CHX Book Feed       & \$0           &       \$0   \\
(Now NYSE Chicago)          & Historical DoB      &               &             \\
Total                       &                     & \$184,000     &  \$90,100   \\ \hline
\end{tabular}
\caption{
The pricing presented in this table assumes a single consumer with an academic use case aiming to construct a dataset similar to what was used in this analysis.
It is also assumed that non-display fees do not apply.
Historical data costs assume a 12 month period of interest, i.e. calendar year 2016.
Strictly speaking, historical data may sufficient for replicating the analysis presented in this paper, making subscription to live feeds unnecessary.
However, utilizing historical data provided by each exchange excludes the possibility of collecting data from a single point of observation, reintroducing the issues of clock synchronization and relativity.
Additionally, highlighting the monthly cost for comprehensive direct feed access shines a light on one of the reasons for the lack of academic participation in the analysis of modern U.S. stock markets.
This does not include costs which may be incurred while curating the data, fulfilling potential co-location requirements, ISP/networking costs, computing hardware acquisition and maintenance, etc.
DoB indicates that a product contains full Depth of Book information (adds, mod, and cancel messages), while ToB indicates that a product contains only Top of Book information (trade and quote messages).
The NYSE Historical ToB product, also called NYSE Daily TAQ, is frequently used in academic studies due to it's relatively low cost and broad coverage (e.g. \cite{bartlett2019rigged} use this product).
Historical data from CHX is not directly available, and the live feeds are transitioning to NYSE technology, thus historical CHX data must be purchased from a third party.
This list is not guaranteed to be comprehensive, additional fees/costs may exist.
*Access to UTP data and NASDAQ direct feed data may granted freely to academic institutions, see UTP Feed Pricing and NASDAQ Academic Waiver Policy for more info.
**Historical data purchased from NYSE only covers 5/21/2018 - present for NYSE National, thus an alternative data provider is required in order to obtain historical data from 2016.
The sources used to construct this table include \href{https://www.ctaplan.com/pricing}{CTA feed pricing}, \href{http://www.utpplan.com/data_admin}{UTP feed pricing} via the Data Policies document, \href{https://www.nyse.com/publicdocs/nyse/data/NYSE_Market_Data_Pricing.pdf}{NYSE feed pricing}, \href{https://www.nyse.com/publicdocs/nyse/data/NYSE_Historical_Market_Data_Pricing.pdf}{NYSE historical data pricing}, \href{http://www.nasdaqtrader.com/TraderB.aspx?id=MDDPricingALLN}{NASDAQ feed pricing}, \href{https://markets.cboe.com/us/equities/membership/fee_schedule/bzx/}{BATS/DirectEdge feed pricing}, and \href{https://www.chx.com/chxshare/fee-schedule.html}{CHX feed pricing}
\label{table:feed pricing}
}
\end{table*}

\section{Calculating realized opportunity cost}
\label{sec:calc-roc}
Calculating Realized Opportunity Cost (ROC)
For each trade of interest:
Obtain the Securities Information Processor (SIP) National Best Bid and Offer (NBBO) prices and the Direct Best Bid and Offer (DBBO) at the time of the trade.
Check if the trade executed at one of the NBBO prices.
If yes, then the difference between the execution price and the corresponding price from the DBBO, multiplied by the number of shares transacted, becomes the ROC associated with that trade.
Note: Depending on the side of the active order (bid or offer), and the relationship between the NBBO and DBBO, the ROC may be identified as favoring the SIP or a Direct feed. In other words, when the active order could receive price improvement by executing at the price displayed by the DBBO, then the ROC becomes associated with the SIP (SIP ROC). Likewise, if the active order received a price improvement by executing at the NBBO rather than the DBBO, then the ROC becomes associated with the Direct feeds (Direct ROC).
If no, then the trade is discarded from the analysis, since it is difficult to accurately determine the side of the active order in this situation—and knowing the side of the active order is required in order to accurately calculate the directional ROC.
Note: ROC experienced on both sides of the book (bid and offer) are aggregated over each day, ticker, and exchange; thus, there may be some cancellation between positive ROC (Direct ROC) and negative ROC (SIP ROC) during the aggregation to determine net ROC for that day-ticker-exchange.  The net ROC is therefore a conservative measure, since it is possible that investors could experience both SIP and Direct ROC for that day-ticker-exchange.

Example:
In particular, see the 79th trade in Table \ref{table:trades_roc}, where 100 shares of AAPL transacted at \$99.13 at 9:48:55.398386. The NBBO at that time was (bid @ \$99.13, offer @ \$99.15), while the DBBO was (bid @ \$99.16, offer @ \$99.17). Since the trade executed at \$99.13, the best bid displayed by the SIP, we infer that the resting order was a bid and the active order was an offer. The ROC is then calculated as (\$99.13 per share - \$99.16 per share) * 100 shares = (-\$0.03 per share) * 100 shares = -\$3.00 in favor of the Direct feeds (i.e. SIP ROC).
From this example, one can note that when the active order is an offer, then the formula for ROC is (SIP National Best Bid (NBB) - Direct Best Bid (DBB)) * shares. This results in a positive value when the NBO provides price improvement for the active bid and a negative value when the DBO provides price improvement for the active bid.
Additionally, see the 95th trade in Table \ref{table:trades_roc} where 100 shares transacted at \$99.14 at 9:48:55.398560. The NBBO at that time was (bid @ \$99.14, offer @ \$99.14) and the DBBO was (bid @ \$99.16, offer @ \$99.17). Since the SIP was locked at the time of execution the active order could have been from either side of the book. For this example, we will focus on the situation where we assume the active order is a bid and the resting order is an offer. The ROC is then calculated as (\$99.17 per share - \$99.14 per share) * 100 shares = (\$0.03 per share) * 100 shares = \$3.00 in favor of the SIP (i.e. Direct ROC).
Note that in this example, the formula used to calculate the ROC reverses the position of the SIP and Direct prices since the active order is a bid instead of an offer. Thus, the formula for ROC is (Direct Best Offer (DBO) - SIP Best Offer (NBO)) * shares.
This maintains the meaning of the sign, where positive values indicate price improvement featured by the NBB and negative values indicate price improvement featured by the DBB (from the perspective of the active order).
Thus, ROC from both sides of the book may be treated uniformly in that positive values favor the SIP feed and negative values favor the consolidated Direct feeds.
We aggregate the ROC by date, stock, and venue.
Since these two trades occurred at the same trading venue, they would be summed, resulting in a net ROC of \$0.00.
Similar cancellations occur for every date-stock-venue combination resulting in these conservative measures of ROC. 
Additionally, in the example dislocation there were almost 100 differing trades (i.e., trades that occurred while the NBBO and DBBO are dislocated) as contained in Table \ref{table:trades_roc}.
Yet, our ROC measures only include trade executions at the NBBO.
Therefore, we only consider a total of 11 trades (6 on the offer and 5 on the bid) during this dislocation, thus providing additional evidence that our ROC measures are conservative.

Dislocations and Dislocation Segments:
In Fig.\ \ref{fig:circle-aapl-1day-real}, we see all dislocations in AAPL on January 7, 2016.  We select an arbitrary dislocation to investigate which existed on the offer side from 9:48:55.396886 to 9:48:55.398749 (a duration of 1863 microseconds).
This dislocation features a maximum value of \$0.06, which occurs between 9:48:55.397644 and 9:48:55.398027 (a duration of 383 microseconds or 20.56\% of its lifetime).
During this time where the dislocation featured its maximum value, the SIP best offer remained at \$99.11 and the Direct best offer remained at \$99.17.
Thus, any bid orders submitted during this period stood to save \$0.06 per share by transacting at the SIP BO rather than the Direct BO, assuming that they could actually locate resting offers at \$99.11, either in the lit or dark markets.
Note that this dislocation started and ended while the Limit-up Limit-down (LULD) mechanism was in effect (this is engaged at 9:45 each day, following the first 15 minutes of trading), featured a duration longer than 545 microseconds (what we consider to be the minimum duration in order to be actionable) and featured a maximum magnitude greater than \$0.01.
Note: you can find more info on LULD here: http://www.luldplan.com/index.html.

Connecting Realized Opportunity Cost and Dislocation Segments:
The ROC statistic captures events that occurred (i.e. trades) and assigns an opportunity cost to them based on the state of the SIP and Direct feeds at the time of the trade.
Hopefully the above example has illustrated the extreme sparsity of our ROC approach, which only considers trades that execute at either side of the prevailing NBBO, features cancellation effects due to aggregation, and does not consider duration/actionability (e.g., could the agent who entered the active order have reasonably reacted to the state of the two feeds?).
Dislocation segments are constructed to capture the relative states between the NBBO and DBBO through time, observing the dislocations between the two feeds and collecting information about their duration and magnitude.
With our approach, we capture the inefficiencies and opportunity costs that actually occurred (i.e., realized), and what inefficiencies and opportunity costs could have occurred (i.e., dislocation segments).
For illustrative purposes only, if the NBBO and DBBO were tightly synchronized, then the ROC statistic would tend towards \$0.00.
[Note: there are specific policy reasons in Reg. NMS that SIP reporting will always lag reporting on the direct feeds, independent of technological infrastructure].
Thus, constructing DSs so that they only consider the NBBO and DBBO allows us to isolate one component of the ROC statistic and investigate it in greater detail.
Additionally, the ROC statistic does not account for duration/actionability, while DSs allow for such considerations of duration / actionability to be addressed in a simple and direct way.
These two measurements, ROC and DSs, were constructed to investigate similar phenomena from slightly different perspectives to provide complementary and synergistic views of NMS dynamics.

{
\setlength{\extrarowheight}{.125em}
\begin{longtable*}[c]{ | cccccccc | }
\hline
Index & Thesys Timestamp & Delta & Symbol & Size &  Price & Exchange & Extra \\
\hline
\endhead
\hline
\multicolumn{8}{|r|}{{Continued on next page\rule{0pt}{2.6ex}}} \\ \hline
\endfoot

\endlastfoot
0  &  2016-01-07 09:48:55.396951 &    255 &   AAPL &   100 &  99.11 &                1 &   -651 \\
1  &  2016-01-07 09:48:55.396951 &    227 &   AAPL &   100 &  99.12 &                1 &   -651 \\
2  &  2016-01-07 09:48:55.396978 &    237 &   AAPL &   100 &  99.12 &                1 &   -678 \\
3  &  2016-01-07 09:48:55.396978 &    222 &   AAPL &   100 &  99.12 &                1 &   -678 \\
4  &  2016-01-07 09:48:55.396978 &    204 &   AAPL &   100 &  99.13 &                2 &   -852 \\
5  &  2016-01-07 09:48:55.396998 &    207 &   AAPL &   100 &  99.13 &                2 &   -872 \\
6  &  2016-01-07 09:48:55.396998 &    190 &   AAPL &   100 &  99.13 &                2 &   -872 \\
7  &  2016-01-07 09:48:55.397064 &    239 &   AAPL &   100 &  99.13 &                2 &   -938 \\
8  &  2016-01-07 09:48:55.397064 &    216 &   AAPL &   100 &  99.12 &                1 &   -764 \\
9  &  2016-01-07 09:48:55.397068 &    204 &   AAPL &    50 &  99.13 &                2 &   -942 \\
10 &  2016-01-07 09:48:55.397196 &    316 &   AAPL &   200 &  99.13 &                2 &  -1070 \\
11 &  2016-01-07 09:48:55.397196 &    296 &   AAPL &   100 &  99.16 &                1 &  -1013 \\
12 &  2016-01-07 09:48:55.397196 &    279 &   AAPL &   100 &  99.13 &                2 &  -1070 \\
13 &  2016-01-07 09:48:55.397196 &    262 &   AAPL &   395 &  99.11 &                3 &  -1044 \\
14 &  2016-01-07 09:48:55.397297 &    344 &   AAPL &   100 &  99.13 &                1 &   -997 \\
15 &  2016-01-07 09:48:55.397297 &    327 &   AAPL &   100 &  99.16 &                4 &  -1114 \\
16 &  2016-01-07 09:48:55.397297 &    309 &   AAPL &   100 &  99.13 &                1 &   -997 \\
17 &  2016-01-07 09:48:55.397297 &    292 &   AAPL &   100 &  99.14 &                2 &  -1171 \\
18 &  2016-01-07 09:48:55.397297 &    275 &   AAPL &   100 &  99.13 &                1 &   -997 \\
19 &  2016-01-07 09:48:55.397297 &    259 &   AAPL &   100 &  99.12 &                3 &  -1145 \\
20 &  2016-01-07 09:48:55.397361 &    306 &   AAPL &   100 &  99.14 &                2 &  -1235 \\
21 &  2016-01-07 09:48:55.397431 &    358 &   AAPL &   100 &  99.13 &                3 &  -1279 \\
22 &  2016-01-07 09:48:55.397431 &    317 &   AAPL &   100 &  99.14 &                2 &  -1305 \\
23 &  2016-01-07 09:48:55.397431 &    298 &   AAPL &   100 &  99.13 &                1 &  -1131 \\
24 &  2016-01-07 09:48:55.397431 &    268 &   AAPL &   100 &  99.13 &                3 &  -1279 \\
25 &  2016-01-07 09:48:55.397499 &    316 &   AAPL &    50 &  99.13 &                1 &  -1199 \\
26 &  2016-01-07 09:48:55.397499 &    299 &   AAPL &   100 &  99.13 &                3 &  -1347 \\
27 &  2016-01-07 09:48:55.397499 &    284 &   AAPL &   100 &  99.13 &                1 &  -1199 \\
28 &  2016-01-07 09:48:55.397504 &    272 &   AAPL &   100 &  99.14 &                2 &  -1378 \\
29 &  2016-01-07 09:48:55.397504 &    255 &   AAPL &   100 &  99.13 &                1 &  -1204 \\
30 &  2016-01-07 09:48:55.397565 &    299 &   AAPL &    50 &  99.13 &                3 &  -1413 \\
31 &  2016-01-07 09:48:55.397565 &    281 &   AAPL &   100 &  99.14 &                1 &  -1265 \\
32 &  2016-01-07 09:48:55.397565 &    266 &   AAPL &   200 &  99.13 &                3 &  -1413 \\
33 &  2016-01-07 09:48:55.397604 &    290 &   AAPL &   100 &  99.14 &                1 &  -1304 \\
34 &  2016-01-07 09:48:55.397604 &    276 &   AAPL &   100 &  99.13 &                3 &  -1452 \\
35 &  2016-01-07 09:48:55.397604 &    260 &   AAPL &   100 &  99.14 &                1 &  -1304 \\
36 &  2016-01-07 09:48:55.397685 &    325 &   AAPL &   100 &  99.14 &                3 &  -1533 \\
37 &  2016-01-07 09:48:55.397685 &    309 &   AAPL &   100 &  99.14 &                3 &  -1533 \\
38 &  2016-01-07 09:48:55.397685 &    293 &   AAPL &   100 &  99.14 &                1 &  -1385 \\
39 &  2016-01-07 09:48:55.397731 &    323 &   AAPL &   100 &  99.14 &                3 &  -1579 \\
40 &  2016-01-07 09:48:55.397731 &    309 &   AAPL &   100 &  99.14 &                1 &  -1431 \\
41 &  2016-01-07 09:48:55.397731 &    294 &   AAPL &   100 &  99.14 &                3 &  -1579 \\
42 &  2016-01-07 09:48:55.397731 &    279 &   AAPL &    50 &  99.15 &                1 &  -1431 \\
43 &  2016-01-07 09:48:55.397767 &    300 &   AAPL &   100 &  99.14 &                3 &  -1615 \\
44 &  2016-01-07 09:48:55.397767 &    285 &   AAPL &   100 &  99.14 &                3 &  -1615 \\
45 &  2016-01-07 09:48:55.397767 &    269 &   AAPL &   900 &  99.15 &                3 &  -1615 \\
46 &  2016-01-07 09:48:55.397824 &    310 &   AAPL &   100 &  99.15 &                1 &  -1524 \\
47 &  2016-01-07 09:48:55.397824 &    294 &   AAPL &   100 &  99.15 &                3 &  -1672 \\
48 &  2016-01-07 09:48:55.397824 &    280 &   AAPL &   100 &  99.14 &                2 &  -1698 \\
49 &  2016-01-07 09:48:55.397824 &    266 &   AAPL &   100 &  99.15 &                2 &  -1698 \\
50 &  2016-01-07 09:48:55.397870 &    298 &   AAPL &   100 &  99.15 &                2 &  -1744 \\
51 &  2016-01-07 09:48:55.397870 &    282 &   AAPL &   100 &  99.15 &                1 &  -1570 \\
52 &  2016-01-07 09:48:55.397894 &    290 &   AAPL &   100 &  99.15 &                3 &  -1742 \\
53 &  2016-01-07 09:48:55.397894 &    275 &   AAPL &   100 &  99.15 &                1 &  -1594 \\
54 &  2016-01-07 09:48:55.397894 &    260 &   AAPL &    50 &  99.15 &                3 &  -1742 \\
55 &  2016-01-07 09:48:55.397973 &    323 &   AAPL &    50 &  99.15 &                1 &  -1673 \\
56 &  2016-01-07 09:48:55.397973 &    307 &   AAPL &   100 &  99.15 &                3 &  -1821 \\
57 &  2016-01-07 09:48:55.397973 &    293 &   AAPL &   100 &  99.15 &                1 &  -1673 \\
58 &  2016-01-07 09:48:55.397994 &    299 &   AAPL &    50 &  99.15 &                2 &  -1868 \\
59 &  2016-01-07 09:48:55.398058 &    346 &   AAPL &    50 &  99.16 &                1 &  -1758 \\
60 &  2016-01-07 09:48:55.398058 &    331 &   AAPL &   200 &  99.15 &                3 &  -1906 \\
61 &  2016-01-07 09:48:55.398058 &    313 &   AAPL &   100 &  99.16 &                1 &  -1758 \\
62 &  2016-01-07 09:48:55.398125 &    366 &   AAPL &   100 &  99.15 &                2 &  -1999 \\
63 &  2016-01-07 09:48:55.398128 &    354 &   AAPL &   100 &  99.15 &                3 &  -1976 \\
64 &  2016-01-07 09:48:55.398147 &    357 &   AAPL &   100 &  99.14 &                1 &  -1422 \\
65 &  2016-01-07 09:48:55.398147 &    342 &   AAPL &   200 &  99.15 &                2 &  -2021 \\
66 &  2016-01-07 09:48:55.398158 &    339 &   AAPL &   100 &  99.16 &                3 &  -2006 \\
67 &  2016-01-07 09:48:55.398177 &    342 &   AAPL &   100 &  99.15 &                2 &  -2051 \\
68 &  2016-01-07 09:48:55.398225 &    375 &   AAPL &   100 &  99.16 &                3 &  -2073 \\
69 &  2016-01-07 09:48:55.398225 &    359 &   AAPL &     5 &  99.16 &                3 &  -2073 \\
70 &  2016-01-07 09:48:55.398225 &    345 &   AAPL &   100 &  99.14 &                1 &  -1500 \\
71 &  2016-01-07 09:48:55.398267 &    373 &   AAPL &   100 &  99.15 &                1 &  -1542 \\
72 &  2016-01-07 09:48:55.398267 &    358 &   AAPL &   100 &  99.16 &                1 &  -1542 \\
73 &  2016-01-07 09:48:55.398267 &    342 &   AAPL &   100 &  99.17 &                1 &  -1542 \\
74 &  2016-01-07 09:48:55.398267 &    327 &   AAPL &   100 &  99.17 &                1 &  -1542 \\
75 &  2016-01-07 09:48:55.398267 &    312 &   AAPL &    50 &  99.17 &                1 &  -1542 \\
76 &  2016-01-07 09:48:55.398272 &    300 &   AAPL &   400 &  99.11 &                5 &  -1967 \\
77 &  2016-01-07 09:48:55.398272 &    285 &   AAPL &   100 &  99.12 &                5 &  -1967 \\
78 &  2016-01-07 09:48:55.398386 &    384 &   AAPL &   100 &  99.13 &                5 &  -2081 \\
79 &  2016-01-07 09:48:55.398414 &    397 &   AAPL &   100 &  99.13 &                5 &  -2109 \\
80 &  2016-01-07 09:48:55.398414 &    381 &   AAPL &   100 &  99.13 &                5 &  -2109 \\
81 &  2016-01-07 09:48:55.398414 &    365 &   AAPL &   100 &  99.13 &                5 &  -2109 \\
82 &  2016-01-07 09:48:55.398444 &    381 &   AAPL &    50 &  99.13 &                5 &  -2139 \\
83 &  2016-01-07 09:48:55.398444 &    366 &   AAPL &   100 &  99.13 &                5 &  -2139 \\
84 &  2016-01-07 09:48:55.398444 &    352 &   AAPL &   100 &  99.13 &                5 &  -2139 \\
85 &  2016-01-07 09:48:55.398444 &    337 &   AAPL &   100 &  99.13 &                5 &  -2139 \\
86 &  2016-01-07 09:48:55.398444 &    322 &   AAPL &   100 &  99.14 &                5 &  -2139 \\
87 &  2016-01-07 09:48:55.398532 &    395 &   AAPL &    50 &  99.14 &                5 &  -2227 \\
88 &  2016-01-07 09:48:55.398532 &    369 &   AAPL &   100 &  99.15 &                2 &  -1507 \\
89 &  2016-01-07 09:48:55.398532 &    354 &   AAPL &   100 &  99.16 &                2 &  -1507 \\
90 &  2016-01-07 09:48:55.398537 &    344 &   AAPL &    50 &  99.17 &                2 &  -1512 \\
91 &  2016-01-07 09:48:55.398537 &    330 &   AAPL &   100 &  99.17 &                2 &  -1512 \\
92 &  2016-01-07 09:48:55.398560 &    339 &   AAPL &   100 &  99.17 &                2 &  -1535 \\
93 &  2016-01-07 09:48:55.398560 &    324 &   AAPL &    50 &  99.17 &                3 &  -1282 \\
94 &  2016-01-07 09:48:55.398560 &    309 &   AAPL &   100 &  99.14 &                5 &  -1434 \\
95 &  2016-01-07 09:48:55.398571 &    305 &   AAPL &    50 &  99.15 &                5 &  -1445 \\
96 &  2016-01-07 09:48:55.398571 &    291 &   AAPL &   100 &  99.15 &                5 &  -1445 \\
\hline
\noalign{\vskip 1mm}
\caption{
Trades that occurred during a dislocation in AAPL on 2016-01-07 at approximately 9:48am, more than three minutes after the trading ``guardrails'' are enforced.
The ``Delta" column indicates the difference between the Thesys timestamp and the SIP publication timestamp (in microseconds).
For trade 0, Thesys received the trade at 9:48:55.396951 and the SIP timestamp was 9:48:55.396696.
The ``Extra" column contains additional deltas related to the timestamps added in the 2015 SIP changes, see \cite{bartlett2019rigged} for additional details.
In particular, this column contains the difference (in microseconds) between the Thesys timestamp and the exchange timestamp.
For trade 0, Thesys received the trade at 9:48:55.396951 and the exchange timestamp was 9:48:55.397602, an example of the timestamp inversion seen in \cite{bartlett2019rigged}, which is generally cause by clock synchronization issues.
}
\label{table:trades_roc}
\end{longtable*}
}

\begin{table*}
\setlength{\extrarowheight}{.125em}
\begin{tabular}{|lrrrrrrrr|}
\hline
timestamp &  Exchange number &  price &  shares &  direct\_bid &  direct\_ask &  sip\_bid &  sip\_ask &   roc \\

\hline
2016-01-07 09:48:55.396951 &                1 &  99.11 &     100 &       99.14 &       99.14 &    99.10 &    99.11 &   3.0 \\
2016-01-07 09:48:55.397196 &                3 &  99.11 &     395 &       99.14 &       99.15 &    99.10 &    99.11 &  15.8 \\
2016-01-07 09:48:55.398147 &                1 &  99.14 &     100 &       99.16 &       99.17 &    99.12 &    99.14 &   3.0 \\
2016-01-07 09:48:55.398225 &                3 &  99.14 &     100 &       99.16 &       99.17 &    99.12 &    99.14 &   3.0 \\
2016-01-07 09:48:55.398532 &                2 &  99.15 &     100 &       99.16 &       99.17 &    99.14 &    99.15 &   2.0 \\
2016-01-07 09:48:55.398560 &                5 &  99.14 &     100 &       99.16 &       99.17 &    99.14 &    99.14 &   3.0 \\
\hline
\end{tabular}
\caption{
A subset of the trades from Table \ref{table:trades_roc} that resulted in positive ROC.
Positive ROC indicates that these trades received favorable prices that were aligned with the SIP NBBO.
}
\label{tab:ask_roc_trades}
\end{table*}

\begin{table*}
\setlength{\extrarowheight}{.125em}
\begin{tabular}{|lrrrrrrrr|}
\hline
timestamp &  Exchange number &  price &  shares &  direct\_bid &  direct\_ask &  sip\_bid &  sip\_ask &  roc \\
\hline
2016-01-07 09:48:55.398272 &                5 &  99.12 &     100 &       99.16 &       99.17 &    99.12 &    99.14 &   -4 \\
2016-01-07 09:48:55.398386 &                5 &  99.13 &     100 &       99.16 &       99.17 &    99.13 &    99.15 &   -3 \\
2016-01-07 09:48:55.398444 &                5 &  99.14 &     100 &       99.16 &       99.17 &    99.14 &    99.15 &   -2 \\
2016-01-07 09:48:55.398532 &                5 &  99.14 &      50 &       99.16 &       99.17 &    99.14 &    99.15 &   -1 \\
2016-01-07 09:48:55.398560 &                5 &  99.14 &     100 &       99.16 &       99.17 &    99.14 &    99.14 &   -2 \\
\hline
\end{tabular}
\caption{
A subset of the trades from Table \ref{table:trades_roc} that resulted in negative ROC.
Negative ROC indicates that these trades executed at less favorable prices than what was offered by the DBBO.
}
	\label{tab:bid_roc_trades}
\end{table*}


%

\end{document}